\documentclass[a4paper,11pt]{article}
\pdfoutput=1 

\usepackage{jheppub} 
            
\usepackage{bm,amssymb,slashed,graphicx,multirow,soul,mathtools,xspace,array}  
\usepackage{float}                   
\allowdisplaybreaks
\usepackage{ bbold }
\usepackage{subfigure}

\newcommand{\todo}[1]{{\color{red} \ifmmode\else[todo]\fi #1}}
\usepackage[usenames,dvipsnames]{xcolor}
     \definecolor{hgreen}{rgb}{0,.3,0}
      \definecolor{darkgreen}{rgb}{0.3,.8,0.2}
     \definecolor{hred}{rgb}{.3,0,0}
     \definecolor{hblue}{rgb}{0,0,.3}
     \definecolor{LightGray}{gray}{0.95}

\definecolor{sanddune}{rgb}{0.59, 0.44, 0.09}

\usepackage{hyperref}
\usepackage[normalem]{ulem}

\usepackage{braket}
\usepackage{booktabs}
\usepackage{multicol}
\usepackage{tabularx}
\usepackage{multirow}

\definecolor{mypink}{RGB}{219, 48, 122}

\newcommand{\GeV}{{\, \rm GeV}}

\newcommand{\TeV}{{\, \rm TeV}}

\def\beq{\begin{equation}}
\def\eeq{\end{equation}}

\preprint{P3H-22-058, TTP22-036, DO-TH 22/16}

\title{The Charged Higgs from the Bottom-Up: \\ Probing Flavor at the LHC}

\author[a]{Nishita Desai,}
\author[b]{Alberto Mariotti,}
\author[c]{Mustafa Tabet,}
\author[d]{Robert Ziegler,}

\affiliation[a]{Tata Institute of Fundamental Research, Homi Bhabha Road, Mumbai 400005, India}
\affiliation[b]{Theoretische Natuurkunde and IIHE/ELEM, Vrije Universiteit Brussel,
\\ \& The International Solvay Institutes, Pleinlaan 2, B-1050 Brussels, Belgium}
\affiliation[b]{Fakult\"at f\"ur Physik, TU Dortmund, D-44221 Dortmund, Germany}
\affiliation[d]{Institut f\"ur Theoretische Teilchenphysik, Karlsruhe Institute of Technology (KIT), \\ 76128 Karlsruhe, Germany
}

\emailAdd{desai@theory.tifr.res.in}
\emailAdd{alberto.mariotti@vub.be}
\emailAdd{mustafa.tabet@tu-dortmund.de}
\emailAdd{robert.ziegler@kit.edu}

\date{\today}

\abstract{We systematically study model-independent constraints on the three generic charged Higgs couplings to $b$-quarks and  up-type quarks. While existing LHC searches have focussed on the $tb$ coupling, we emphasize that the LHC plays a crucial role in probing  also $ub$ and $cb$ couplings, since constraints from flavor physics are  weak. In particular we propose various new searches that can significantly extend the present reach on the parameter space by: i)~looking for light charged Higgses that decay into $ub$-quarks, ii)~probing charged Higgs couplings to light and top quarks using multi-$b$-jet signatures, iii)~looking for single $b$-quarks in low-mass dijet searches,  iv)~searching for charge asymmetries induced by charged Higgs production via $ub$ couplings.  }

\begin{document}

 \maketitle

\newpage

\section{Introduction}

Extended Higgs sectors are ubiquitous in theories beyond the Standard Model (SM), and might play an important role to solve long-standing problems in particle physics, such as the stability of the electroweak scale~\cite{Haber:1984rc}, the nature of Dark Matter (DM)~\cite{Deshpande:1977rw}, the absence of CP violation in strong interactions~\cite{DFSZ1, DFSZ2}, or the origin of the matter/anti-matter asymmetry~\cite{Kuzmin:1985mm}. Accordingly, the experimental collaborations have been actively searching for  additional Higgs bosons at colliders, resulting in stringent constraints on common benchmark models, such as  the so-called Type I and II two Higgs doublet models (2HDM), which by construction avoid the appearance of large flavor-changing neutral currents~\cite{Glashow:1976nt}. However, it is fairly easy to evade the standard collider bounds on additional neutral and charged Higgs bosons by allowing for sizable flavor-changing interactions~\cite{Gori:2017tvg}, which at the same time are consistent with constraints from precision flavor physics. Such interactions  are present in the very general class of Type-III 2HDMs, and are controlled by a plethora of free parameters that gives rise to an extremely rich phenomenology~\cite{Atwood:1996vj,Crivellin:2013wna}. It is essential to be aware of these possibilities when designing experimental search strategies aiming to cover the blind spots of traditional searches. 

Specific pattern of flavor violation have been employed in numerous scenarios motivated by outstanding theoretical problems and/or experimental anomalies. Flavor-changing couplings of light additional Higgs fields have been considered for example in QCD axion models (accounting for both Strong CP and  DM)~\cite{Chiang:2015cba,Badziak:2021apn}, scenarios of Electroweak Baryogenesis~\cite{Chiang:2016vgf, Fuyuto:2017ewj, Modak:2020uyq, Hou:2020tnc}, and models addressing fermion mass hierarchies~\cite{Bauer:2015kzy,Bauer:2015fxa, Altmannshofer:2015esa, Ghosh:2015gpa, Altmannshofer:2016zrn, Dery:2016fyj} or the origin of CP violation in the Yukawa sector~\cite{Lee:1973iz, Nierste:2019fbx}. Similar setups have been used to explain the anomalous magnetic moment of the muon~ \cite{Omura:2015nja, Arroyo-Urena:2015uoa, Li:2018aov} or the various anomalies observed in semi-leptonic $B$-meson decays~\cite{Celis:2012dk,Crivellin:2012ye,Kim:2015zla,Iguro:2017ysu,Crivellin:2015mga, Iguro:2018qzf, Li:2018rax}. Phenomenological aspects of these scenarios have been studied in e.g. Refs.~\cite{Cheng:1987rs,Hou:1991un,Branco:1996bq, Pich:2009sp, Altmannshofer:2012ar,Altunkaynak:2015twa,Arroyo-Urena:2013cyf,Botella:2015hoa,Bertuzzo:2015ada,Gori:2017qwg, Crivellin:2017upt, Kohda:2017fkn, Altmannshofer:2018bch, Altmannshofer:2019ogm, Ghosh:2019exx, Hou:2020chc, Hou:2021xiq, Bordone:2021cca, Iguro:2022uzz, Blanke:2022pjy}. Given the broad spectrum of theoretically motivated scenarios and phenomenological implications, it is appropriate to use a model-independent approach focussing on a specific sector of flavor-violating Higgs couplings, in the same spirit as e.g. Ref.~\cite{Gori:2017tvg}. 

In the present work we analyze the present experimental constraints and future prospects on generic couplings of a charged Higgs boson to $b$-quarks, described  by the simplified Lagrangian 
\begin{align}
{\cal L} =  H^+ \left( g_{t b} \overline{t}_R b_L +  g_{c b} \overline{c}_R b_L +  g_{u b} \overline{u}_R b_L \right) + {\rm h.c.} - m_{H^\pm}^2 H^+ H^- \, .
\label{Lag}
\end{align}
We neglect $H^+$ couplings to right-handed $b$-quarks because they are connected to flavor-violating couplings of neutral Higgses to down-type quarks, which are strongly constrained by flavor physics\footnote{The couplings in Eq.~\eqref{Lag} instead are connected to flavor-violating couplings of neutral Higgses to up-type quarks, which are only weakly constrained by flavor observables.}. We also ignore charged Higgs couplings to leptons, which could play an important role in scenarios explaining the current anomalies in $R_{D^{(*)}}$ measured at the $B$-factories (see e.g. Ref.~\cite{Iguro:2022uzz}). Here instead we focus on the case where such couplings are sufficiently small such that the charged Higgs decays dominantly into quarks. A particularly motivated UV scenario behind the Lagrangian in Eq.~\eqref{Lag} is provided by the class of models in Ref.~\cite{Nierste:2019fbx}, which induces the CKM phase by spontaneous CP violation at the electroweak scale. Interestingly, this implies both an upper bound on the charged Higgs mass of about 430 GeV and a lower bound on its couplings to $b$-quarks, ${\rm max} (|g_{t b}|,|g_{c b}|,|g_{u b}|) \ge 0.20$.

The couplings in Eq.~\eqref{Lag} are only weakly constrained by flavor physics, as $D$--$\overline{D}$ mixing only probes the combination $g_{ub} g_{cb}$ and top decays such as $t \to u \gamma,c \gamma$ give negligible constraints.  Therefore bounds from direct searches at colliders are crucial in order to probe charged Higgs couplings to $b$-quarks. In the following we study these constraints using (and recasting) existing searches at the LHC. They can be classified according to the  production and decay topology\footnote{We classify the topologies on the quark level, which is just a shorthand for the corresponding LHC process, e.g. $qb  \to H^\pm \to  qb$ stands for $ p p \to (j) H^\pm (\to qb) $, where $(j)$ denotes possible extra jets, while $tb  \to H^\pm \to  tb$ stands for $p p \to tb H^\pm (\to tb)$.  }: 
\begin{align*}
& {\rm {\bf 1)} \, \,  top \, associated \, production \, and \, top \, decay \, } &  
tb & \to H^\pm \to  tb \, , \\
& {\rm {\bf 2)} \, \, top \, associated  \, production \, and \, light \, quark \, decay \, } &  
 tb & \to H^\pm \to  qb  \, , \\
& {\rm {\bf 3)} \, \, light \, quark \, production \, and \, decay \, } &  
 qb & \to H^\pm \to  qb \, , \\
& {\rm {\bf 4)}  \, \, light \, quark \, production \, and \, top \, decay \, } &
qb & \to H^\pm  \to  tb \, ,
\end{align*}
where $q = u,c$, since it is typically difficult to distinguish up and charm quarks~\cite{ATLAS:2021cxe, CMS:2021scf, CMS:2022psv}. 

Out of these general topologies, at present only top production and decay has been studied by the experimental collaborations~\cite{tbtb_ATLAS,tbtb_CMS}, in addition to top production and $cb$ decay when  the decay to $tb$ is kinematically forbidden~\cite{tbcb_CMS}. 
Signal topologies with only light quarks have not been explicitly searched for, but can be constrained by  recasting dijet constraints on $Z^\prime$ models~\cite{ATLAS:2018qto,CMS:2018mgb, Aaboud:2019zxd, CMS:2019emo,CMS:2019mcu} (see also Ref.~\cite{Bordone:2021cca}). In this work we systematically include all available channels to derive constraints on the generic couplings in Eq.~\eqref{Lag}. Moreover, we suggest to extend existing experimental searches in several ways: {\bf 1)} we propose to carry out a search for $tb \to H^\pm \to ub$ in a similar way to the existing search for $tb \to H^\pm \to cb$,  {\bf 2)} we suggest to perform dedicate searches for charged Higgses coupling dominantly to light quarks, $qb \to H^\pm \to qb$, using single $b$-quark tagging,   {\bf 3)} we study the potential to test the top decay channels  $qb \to H^\pm \to tb$ using multi-$b$-jet signatures (as has been proposed also in Ref.~\cite{Ghosh:2019exx}), and finally  {\bf 4)} we  study the possibility to distinguish $c$- and $u$-quarks in $qb \to H^\pm \to tb$ using charge asymmetries. 

This work is structured as follows: In Section~\ref{Flavor} we  study the flavor constraints from $D\text{--}\overline{D}$ mixing and top decays. In Section~\ref{Collider} we discuss the existing collider constraints for the four production and decay topologies above, and propose various new searches that could significantly increase sensitivity. In particular we study the potential to use lepton charge asymetries in order to distinguish $ub$ from $cb$ production and reduce background. In Section~\ref{Combined constraints} we express the constraints on the various signal topologies at colliders as model-independent bounds on the Lagrangian couplings in Eq.~\eqref{Lag}, also combining with the constraints from flavor physics. We summarize our results in Section~\ref{Conclusion}. In Appendix~\ref{validation1} and \ref{validation2} we provide validations of the \texttt{PYTHIA} code used in our analyses for $qb \to H^\pm \to qb$ and $qb \to H^\pm \to tb$, respectively, while Appendix~\ref{BDT} contains the details on multivariate analysis for the $qb \to H^\pm \to tb$ based on a boosted-decision-tree algorithm.

\section{Flavor Constraints}
\label{Flavor}
Box diagrams with charged Higgs exchange provide constraints on the combination $g_{u b} g_{cb}$ from $D$--$\overline{D}$ mixing, while radiative top decays $t \to c \gamma$ ($t \to u \gamma$)
constrain the combination $g_{t b} g_{cb}$ ($g_{t b} g_{ub}$). Only the former provide relevant bounds, as we discuss in the following.

\begin{figure}[t]
\centering
    \includegraphics[width=0.4\textwidth]{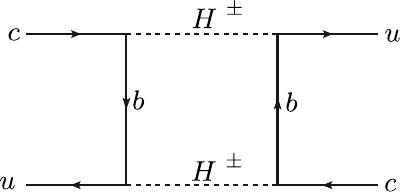}
    \qquad
    \includegraphics[width=0.4\textwidth]{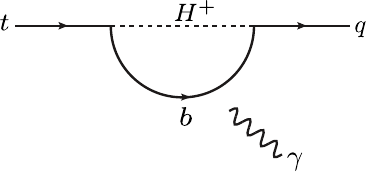}
     \caption{Box diagram contributing to $D$--$\overline{D}$  mixing (left), and photon penguin inducing rare top decays $t\rightarrow q\gamma$, $q=u,c$ (right),  where the photon is attached to any charged particle.}
    \label{fig:flavor diagrams}
\end{figure}

\subsection{$D$--$\overline{D}$ mixing}
The dominant contribution to $D$--$\overline{D}$ mixing comes from
the charged Higgs box diagram in Fig.~\ref{fig:flavor diagrams}, which contributes to the effective $\Delta C= 2$ operator $O^{\prime}_1$.
Assuming that new physics contributions to the absorptive part of the mixing amplitude $\Gamma^{\mathrm{NP}}_{12}$  in the neutral $D$-meson system are negligible, the relevant constraints arise from charged Higgs contributions to the dispersive part of the mixing amplitude $M^\mathrm{NP}_{12} $ 
\begin{align}
    M^\mathrm{NP}_{12} = \frac{1}{2 M_D}\bra{D}H^{\mathrm{NP},|\Delta C| = 2}\ket{\overline{D}}
           =  \frac{\eta \, g_{cb}^2 g^{*2}_{ub}}{128\pi^2 m_{H^\pm}^2}  f \left( \frac{m_b^2}{m_{H^\pm}^2} \right)  \GeV^3 \,,
\end{align}
where $\eta\approx 0.021$ includes the hadronic quantities and the QCD running and the loop function is $f (x) = \left( 1+ 2 x \log x - x^2 \right)/\left( 1- x\right)^3$. 

\begin{figure}[t]
\centering
    \includegraphics[width=0.5\textwidth]{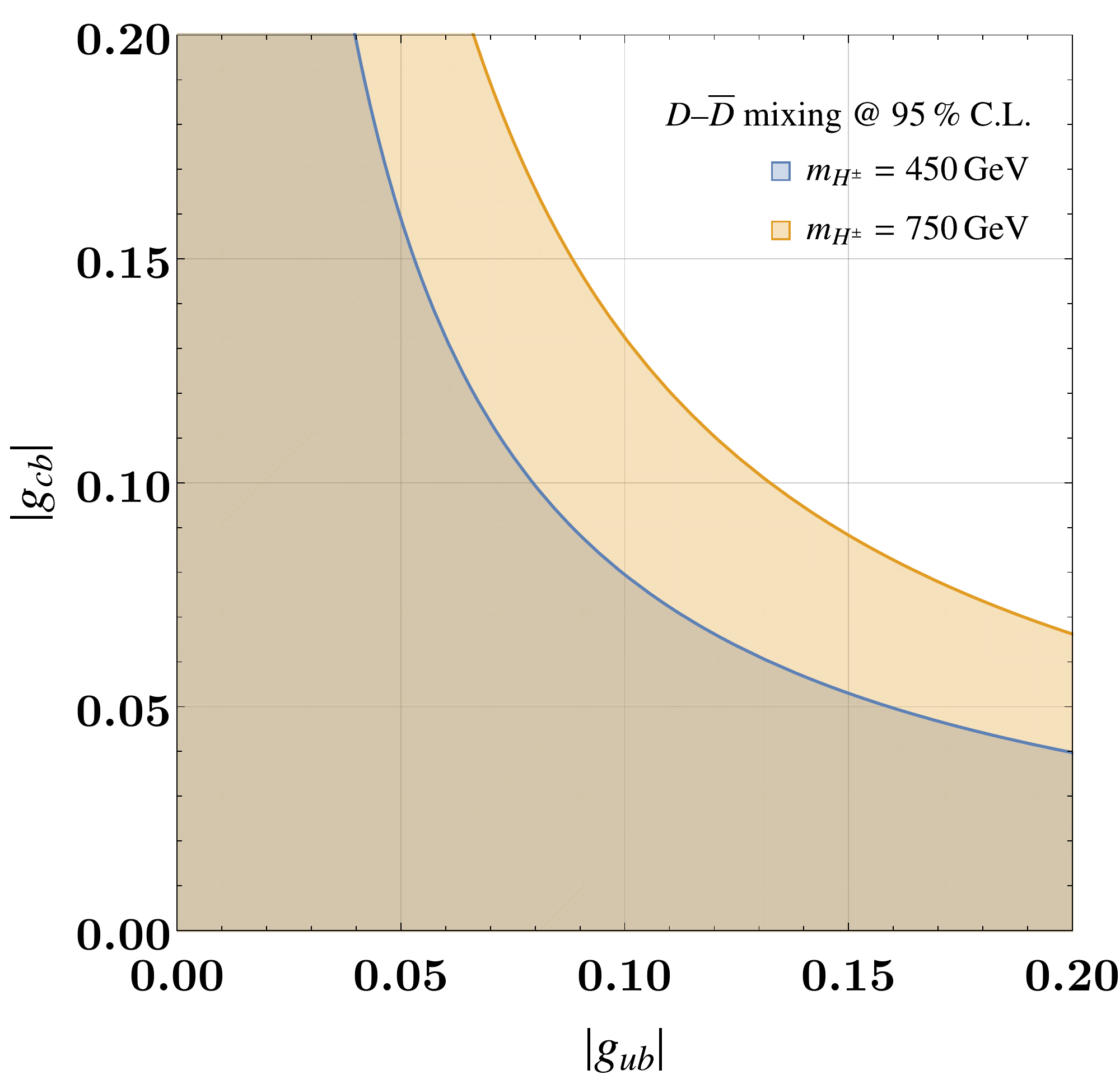}
     \caption{Region in the $g_{ub}$--$g_{cb}$ plane allowed by  $D$-meson mixing at $95\%$ CL. The constraint on the product of both couplings scales linearly with the charged Higgs mass to very good approximation. }
    \label{fig:D-mixing}
\end{figure}

Since the SM contribution is dominated by long-distance
contributions and no reliable estimates are available, we constrain
the charged Higgs contribution to the mixing amplitude 
to lie below the experimental value at $95\%$ CL, which
 is taken from the online update of the UTfit
collaboration~\cite{UTfit:2007eik, UTFIT} 
\begin{align}
|M_{12}|_{\rm 95\%} < 0.0079 \, {\rm ps}^{-1} = 5.2 \times 10^{-15} \GeV \, .
\end{align}
We show the allowed region in the parameter space in Fig.~\ref{fig:D-mixing} for fixed charged Higgs masses, to which the constraints are roughly proportional.  
\subsection{Rare top decays}
The charged Higgs contribution to the rare top decays $t\rightarrow c(u)\gamma$ in Fig.~\ref{fig:flavor diagrams}
gives a constraint on the combination $g_{t b} g_{c(u)b}$.
The current experimental upper limits at $95\%$ CL are~\cite{ATLAS:2019mke}
\begin{align}
    \mathrm{BR}_\mathrm{exp}(t\rightarrow u\gamma) = 6.1\times 10^{-5} \,,\notag\\
    \mathrm{BR}_\mathrm{exp}(t\rightarrow c\gamma) = 1.8\times 10^{-4}  \,.
\end{align}
The charged Higgs contribution is given by
\begin{align}
    \mathrm{BR}(t\rightarrow q\gamma)
    = \frac{m_t^5}{4\pi\Gamma_t}\left(\frac{5e}{1152\pi^2 m_{H^\pm}^2}\right)^2\left|  g_{tb}g^*_{qb} \right|^2 \,,
\end{align}
giving $\mathrm{BR}(t\rightarrow q\gamma)\sim 10^{-6}$
for a charged Higgs mass as light as $100\GeV$ and $\mathcal{O}(1)$ couplings.
Therefore no relevant contraint are currently given by these rare top decays.
\section{Collider Constraints on Production and Decay Topologies}
\label{Collider}
In the following we discuss the collider constraints for all possible production and decay topologies,
assuming 100\% BR in the decay. We will mostly work in the four-flavor-scheme (4FS), except explicitly stated. 

\subsection{Constraints on $tb \to H^\pm \to tb$}
\begin{figure}[t]
\centering
\includegraphics[width=0.35\textwidth]{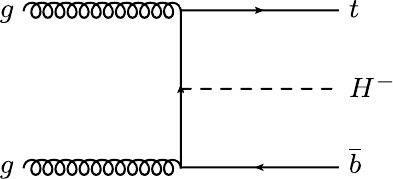}
\qquad
\includegraphics[width=0.35\textwidth]{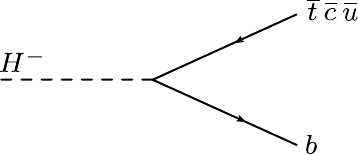}
\caption{Charged Higgs production in association with top and bottom quarks (left) and charged Higgs decay into up-type
and bottom quarks (right).}
\label{fig:tbH}
\end{figure}

\begin{figure}[t]
\centering
    \includegraphics[width=0.6\textwidth]{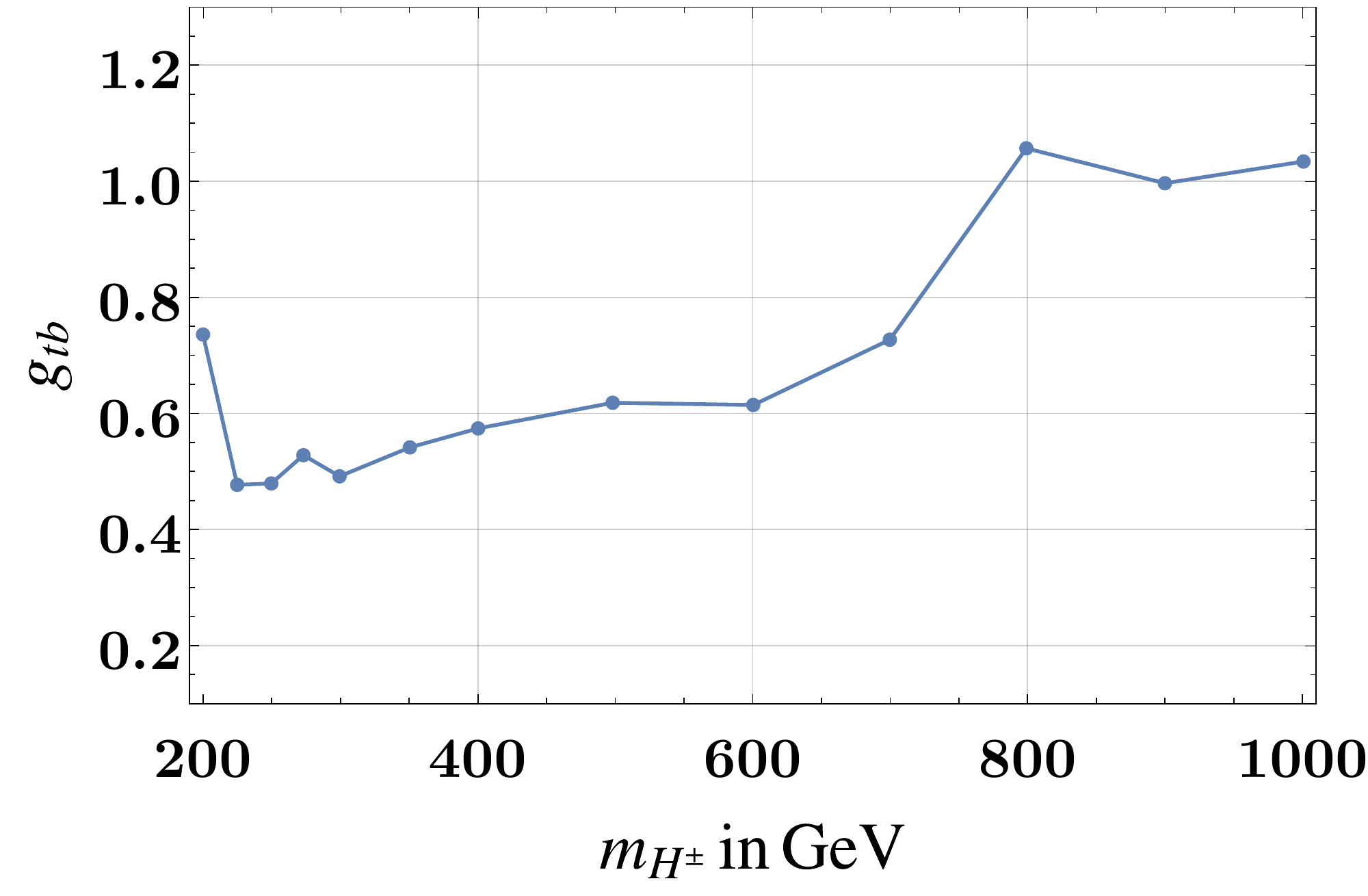}
    \caption{95\% CL Bounds on charged Higgs couplings $g_{tb}$ from top production and decay derived from the ATLAS analysis in Ref.~\cite{tbtb_ATLAS}. 
    \label{fig:tbtb}}
\end{figure}

Searches for charged Higgs bosons that couple mainly to the third generation have been performed both by ATLAS and CMS, looking for multijet events with one lepton and at least 2 $b$-jets The best constraints arise from ATLAS searches at $\sqrt{s} = 13 \TeV$ employing the full Run 2 dataset of $139$\,fb$^{-1}$~\cite{tbtb_ATLAS}, while at present the CMS analysis is based only on $35.9$\,fb$^{-1}$~\cite{tbtb_CMS}. We therefore use only the ATLAS results, which are presented as constraints on the product of the cross section of charged Higgs in association with top and bottom quarks $\sigma (pp \to \overline{t} b H^+)$, see Fig.~\ref{fig:tbH}, and the branching ratio of the charged  Higgs decay BR($H^+ \to t \overline{b}$). We re-interpret this analysis in terms of a constraint on the coupling $g_{tb}$ in Eq.~\eqref{Lag} using the LO cross-sections calculated with \texttt{MadGraph5\_aMC@NLO}~\cite{Alwall:2011uj, Alwall:2014hca} and an approximate K-factor of $K = 1.6$ \cite{Degrande:2016hyf}. The resulting NLO cross-section is used to obtain the 95\% CL bounds on $g_{tb}$ shown in Fig.~\ref{fig:tbtb}. 

\subsection{Constraints on $tb \to H^\pm \to qb$}
\label{tbqb}
If the charged Higgs decays mainly to light quarks, but is dominantly produced via its couplings to the top quark, then there has to be a strong hierarchy $g_{qb} \ll g_{tb}$ and the decay to top quarks must be kinematically closed, so that the charged Higgs is  produced mainly via top decays.  This topology has been searched for both at CMS~\cite{tbcb_CMS} and ATLAS~\cite{tbcb_ATLAS}, giving upper limits on the branching ratio product ${\rm BR}(t \to H^\pm b) \times {\rm BR}(H^\pm  \to c b)$ in the mass range 90--150 GeV (CMS) and 60--160 GeV (CMS). While the CMS search has used $19.7 \, {\rm fb}^{-1}$ of data collected at $\sqrt{s} = 8 \TeV$, the ATLAS analysis is based on a data with $\sqrt{s} = 13 \TeV$ and an integrated luminosity of $139 \, {\rm fb}^{-1}$. Thanks to the larger dataset and refined analysis techniques, the ATLAS search has improved the sensitivity with respect to the CMS analysis by about a factor of five, and also explored a larger $m_{H^\pm}$ range. For this reason we only use the ATLAS results in Ref.~\cite{tbcb_ATLAS}.

This analysis focusses on data enriched in top-quark pair production, where one top quark decays into a leptonically decaying $W$-boson and a bottom quark, and the other top quark decays into a charged Higgs boson and a bottom quark. This topology leads to a lepton-plus-jets final state, characterised by an isolated electron or muon and at least four jets, with a high multiplicity of $b$-jets, and missing energy. A neural network classifier is employed to distinguish between signal and background using kinematic differences. There is an irreducible SM background from $t \overline{t}$ production with a $W$-boson decaying to $cb$, which is however suppressed by the corresponding small CKM element.  

\begin{figure}[t]
\centering
    \includegraphics[width=0.6\textwidth]{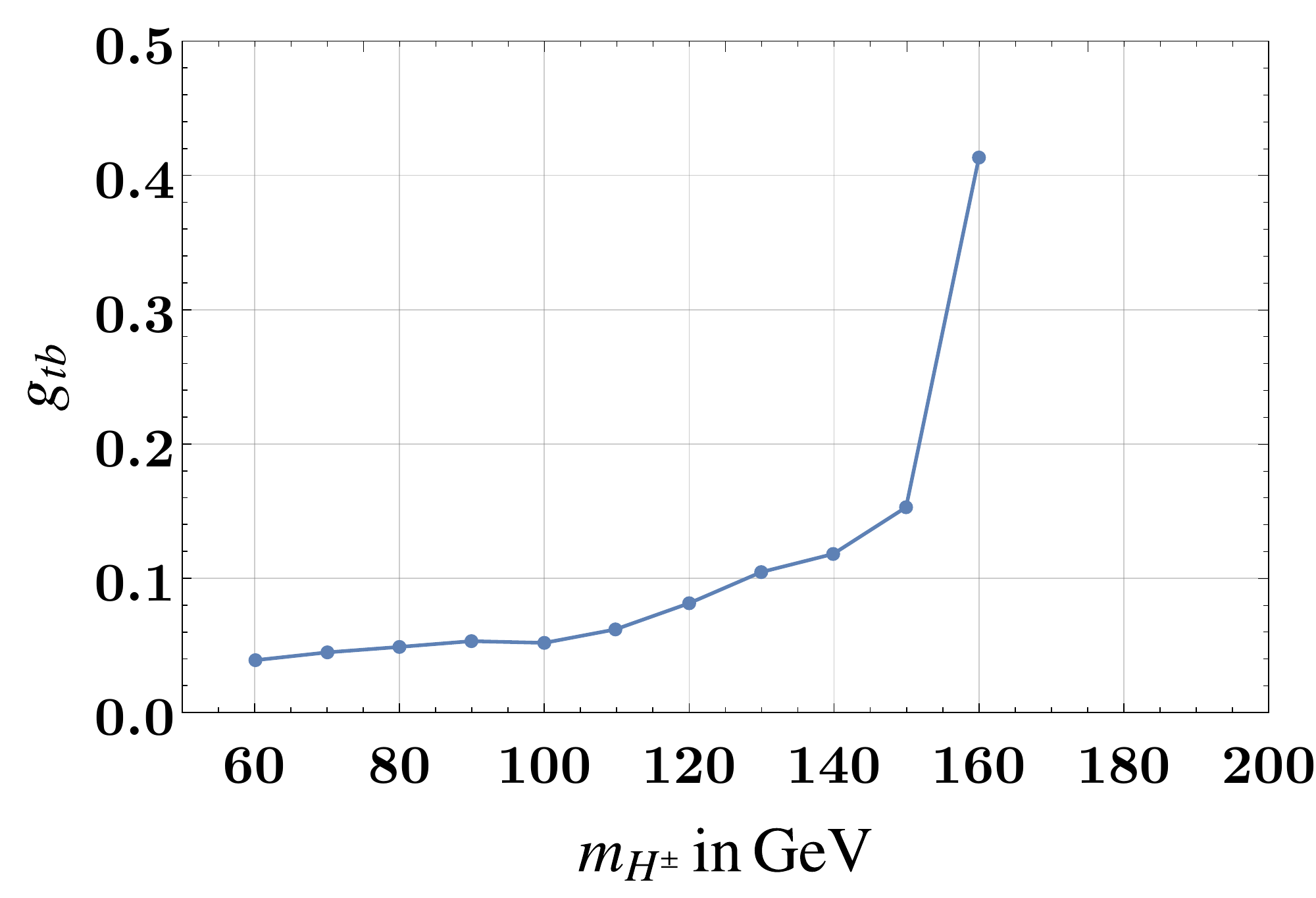}
     \caption{95\% CL constraints on charged Higgs couplings $g_{tb}$ from top production and decay to $cb$, using the ATLAS analysis in Ref.~\cite{tbcb_ATLAS}. These constraints are valid as long as $g_{cb}$ (and $g_{ub}$) are sufficiently small such that $H^\pm$ production is dominated by top decays. 
    \label{fig:tbcb}}
\end{figure}

The resulting ATLAS limits on the branching ratios ${\rm BR}(t \to H^\pm b) \times {\rm BR}(H^\pm  \to c b)$ can be interpreted as upper limits on the top coupling $g_{tb}$, assuming ${\rm BR}(t \to H^\pm b) + {\rm BR}(t \to W^\pm b) = {\rm BR}(H^\pm  \to c b) =1$. These bounds are valid as long as the couplings $g_{ub}, g_{cb}$ are small enough such that the top decays dominate the production. The resulting 95\% CL limits on $g_{tb}$ are shown in Fig.~\ref{fig:tbcb}. 

While the ATLAS analysis in Ref.~\cite{tbcb_ATLAS} has focussed on $H^\pm  \to c b$ decays,  it could be worthwhile to perform a similar analysis on the same data set looking for $H^\pm  \to u b$ decays. We presume such an analysis to be very similar, possibly even using the same signal categories, since the only difference is a light jet instead of a charm-jet. The ATLAS analysis was using a neural network exploiting the $b$-tagging score of this charm-jet, which helps to distinguish the signal from the main backgrounds, $t \overline{t}$ + light jets and  $t \overline{t} + b$-jets. Replacing the charm-jet by a light jet will presumably worsen the separation of the signal from the first background, but improve the separation from the latter background\footnote{We thank Nicola Orlando for clarifying this point.}. Thus one might expect that overall these effects would balance and  the proposed analysis might lead to constraints on ${\rm BR}(t \to H^\pm b) \times {\rm BR}(H^\pm  \to u b)$ that are as stringent than those on ${\rm BR}(t \to H^\pm b) \times {\rm BR}(H^\pm  \to c b)$.

\subsection{Constraints on $qb \to H^\pm \to qb$}
\label{sec:qb qb}
\begin{figure}[t]
\centering
    \includegraphics[width=0.7\textwidth]{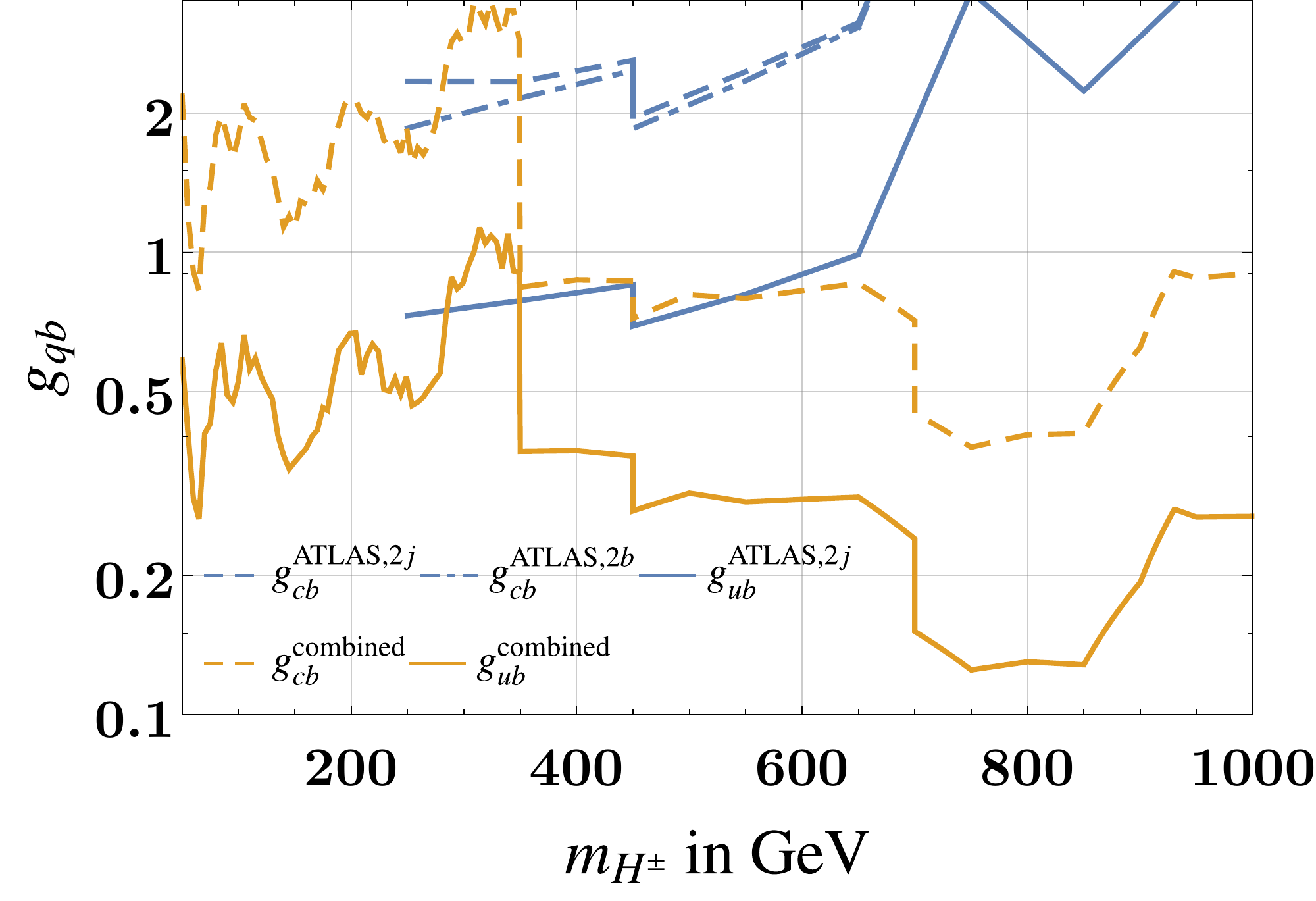}
    \caption{Upper limits at $95\%$ CL on the couplings $g_{ub}$ (solid) and
        $g_{cb}$ (dashed) as a function of the charged-Higgs mass $m_{H^\pm}$ from 
        recasting $Z'$-searches with ISR photons in Ref.~\cite{Aaboud:2019zxd} (blue) and ISR jets Refs.~\cite{CMS:2019emo,ATLAS:2018qto,CMS:2018mgb,CMS:2019mcu} (orange). 
              See text for details.}
    \label{fig:dijetsphoton}
\end{figure}
Dedicated searches for a charged-Higgs boson where light quarks dominate both production
and decay have not been performed by the experimental collaborations. 
However, since the signature of this channel is a dijet resonance, one can put limits on
the couplings by recasting other searches. The most important searches for charged-Higgs with masses of the order of $\sim 300\GeV$ look
for two well separated jets with a high-$p_T$ jet or photon from initial state
radiation (ISR).
We have recasted these searches to obtain the limit shown in Fig.~\ref{fig:dijetsphoton}, which we discuss in more detail in the following.

In Ref.~\cite{Aaboud:2019zxd} a search has been performed for the signature $pp\rightarrow Z^\prime \rightarrow q\overline{q}$ with
an ISR photon. The selected events had been divided into two categories, one where both
jets from the resonant decay are $b$-tagged, and the other without $b$-tagging. Since the resonance does
not decay in two $b$-jets in our case, we focus on the analysis performed for the latter
category\footnote{It would be interesting if the experimental search would also include a \emph{single} $b$-tagged category,
which would presumably give the best sensitivity to our signature as SM background
is further reduced (cf. Ref.~\cite{Iguro:2022uzz}).
For searches that also require an ISR jet, it could be beneficial to additionally require that this ISR jet to be $b$-tagged, in order to reduce the background even further.
}.
However, we use the two $b$-tag category in the case where the resonance decays
predominantly into $cb$, i.e. for a dominant $g_{cb}$ coupling since the
mistag rate of a charm quark is not negligible and might yield sizeable contributions in this
signal region.
Moreover, two different triggers have been used depending on the resonance masses
and differing by the transverse
energy cut on the photon $E^\gamma_{T,\mathrm{trig}}$, and the $p_T$ cut on the jets.
The single-photon trigger, used for resonance masses below $450\GeV$,
requires only a single photon with transverse energy
$E^\gamma_{T,\mathrm{trig}}>150\GeV$.
Above $450\GeV$, the combined trigger is used allowing for a lower transverse energy
cut on the photon $E^\gamma_{T,\mathrm{trig}}>75\ (85)\GeV$ for the $2016\ (2017)$ datasets
by additionally requiring two jet candidates with each $p^\mathrm{jet}_{T,\mathrm{trig}} > 50\GeV$.
Photons are collected in the region $|\eta|<2.37$ excluding $1.37<|\eta|<1.52$.
The selection criteria are listed in Tab.~\ref{tab:selection criteria qbqb}.
Additionally, if a reconstructed jet is not well separated from the isolated
high-$p_T$ photon, i.e. an angular separation of
$\Delta R < 0.4$, the jet candidate is removed.
\begin{table}[t]
\centering
\begin{tabularx}{1\textwidth}{
    >{\raggedright\arraybackslash}p{0.275\textwidth}
    >{\centering\arraybackslash}p{0.33\textwidth}
    >{\centering\arraybackslash}p{0.33\textwidth}}
\toprule
    Criterion & Single-photon trigger & Combined trigger \\
\midrule
    Number of jets          & \multicolumn{2}{c}{$n_\mathrm{jets} \geq 2$} \\
    Number of photons       & \multicolumn{2}{c}{$n_\gamma \geq 1$} \\
    Leading photon          & $E_T^\gamma > 150\GeV$ & $E_T^\gamma >  95~\GeV$ \\
    Leading, subleading jet & $p_T^\mathrm{jet} > 25\GeV$ & $p_T^\mathrm{jet}> 65\GeV$ \\
    Centrality              & \multicolumn{2}{c}{$|y^*|=|y_{1} - y_{2}|/2 < 0.75$}  \\
    Invariant mass          & $m_{jj} > 169\GeV$ & $m_{jj} > 335\GeV$ \\
    Jet $|\eta|$            & \multicolumn{2}{c}{ $|\eta^\mathrm{jet}| < 2.8$} \\
\bottomrule
\end{tabularx}
\caption{Selection criteria of Ref.~\cite{Aaboud:2019zxd}. Here $y_1$ and $y_2$ denote
    the rapidities of the leading and subleading jet (in $p_T$). See text for the other definitions. \label{tab:selection criteria qbqb}}
\end{table}

We generate $pp\rightarrow \gamma H^\pm (\rightarrow qb)$ in the five-flavor-scheme (5FS) in
\texttt{MadGraph5\_aMC@NLO} at leading order using the 2HDM model file given in Ref.~\cite{Degrande:2014vpa}.
To populate the phase space with sufficient events fulfilling the $E^\gamma_T$ cut
on the photons, the events are generated with a $p^\gamma_T$ cut of $p^\gamma_T>100\ (50)\GeV$
for the single-photon (combined) trigger. The analysis is performed in \texttt{PYTHIA 8.2}~\cite{Sjostrand:2006za,Sjostrand:2014zea},
and we have validated our code by applying it to the signature $pp\rightarrow \gamma Z^\prime (\rightarrow q\overline{q})$ and comparing to the
analysis by the ATLAS collaboration in Ref.~\cite{Aaboud:2019zxd}, finding excellent agreement~(cf. Appendix~\ref{validation1}).
To set the limits on the charged-Higgs couplings, we calculate the local significance $Z \approx S/\sqrt{S + B}$
around the resonance, using the observed data from Ref.~\cite{Aaboud:2019zxd} provided
in the online repository HEPData, with the number of signal events $S$ and background events $B$. The resulting upper limits on the charged-Higgs couplings at $95\%$ CL are
shown in Fig.~\ref{fig:dijetsphoton} in blue for $g_{ub}$ (solid) and $g_{cb}$ (dashed).

As it can be seen from this figure, the ISR photon search yields the most stringent constraints
in the mass window between 300 and 350$\GeV$.
Interestingly, the two $b$-tag category yields slightly stronger bounds for
$g_{cb}$ than the category without $b$-tagging, due to the moderate mistag rate of charm quarks.
Below 300$\GeV$ the strongest bounds come instead
from a dijet search with an ISR jet where the two jets stemming from the resonance decay
have been reconstructed as one large-radius jet~\cite{CMS:2019emo}.
Above 350$\GeV$ the experiments usually look for two well-separated jets where the strongest constraints for masses up to
$450\GeV$ again come from searches with an ISR jet~\cite{CMS:2019mcu}.
For even higher masses the generic dijet searches yield the strongest bounds~\cite{ATLAS:2018qto,CMS:2018mgb},
where for masses $\lesssim 1\TeV$ only partial event informations are usually collected in order to not
saturate the trigger.
The ATLAS search~\cite{ATLAS:2018qto} gives the strongest limits for masses below $\approx 930\GeV$, while
the CMS search~\cite{CMS:2018mgb} yields stronger limits for even higher masses. 

In Fig.~\ref{fig:dijetsphoton} we have combined all these constraints and show the strongest one in orange, which is  obtained by
recasting the experimental searches for a ``leptophobic'' $Z^\prime$ vector boson coupling with
universal couplings to quarks in Refs.~\cite{CMS:2019emo,ATLAS:2018qto,CMS:2018mgb,CMS:2019mcu}.
For large resonance masses $> 450$\,GeV, we only apply the appropriate rescaling with the parton
distribution functions (PDFs). For low mass resonances we also include the efficiency ratios
for $Z'$- and $H^\pm$-mediated events, respectively, since here deviations from
the PDF-rescaling can be large due to the presence of an additional ISR jet
yielding sizeable contributions from $gq$-fusion, see e.g. Ref.~\cite{Bordone:2021cca}.

\subsection{Constraints on $qb \to H^\pm \to tb$}
\label{qbtb}
\begin{figure}[t]
\centering
    \includegraphics[width=0.35\textwidth]{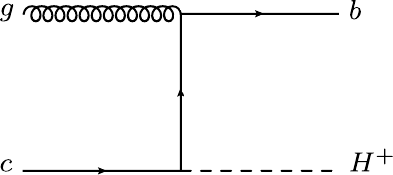}
    \caption{Charged Higgs production through $gc$ fusion contributing to $pp\rightarrow bH^\pm$.}
    \label{fig:gc-fusion}
\end{figure}

The associated production of a charged-Higgs with a single $b$-quark (see Fig.~\ref{fig:gc-fusion}) 
with the subsequent
decay $H^\pm \rightarrow tb$  has not been considered by the experimental collaborations,
so currently there is no dedicated analysis.
The authors of Ref.~\cite{Ghosh:2019exx} have proposed a search for the $cb \to H^\pm \to tb$ topology based on a signature with three $b$-tagged jets plus lepton, which is similar to the searches in Refs.~\cite{tbtb_ATLAS, tbtb_CMS}.
In the following we will revisit the proposed analysis\footnote{We thank the authors of Ref.~\cite{Ghosh:2019exx} for correspondence about their results.}, and extend it by 
{\bf i)} studying also the case of $ub$ production,
{\bf ii)} including the impact of systematic uncertainties and 
{\bf iii)}  exploring a more refined cut strategy (also employing a BDT algorithm) to further optimize the expected sensitivity.

Before explaining the details of the proposed search, 
we note that 
possible constraints on this topology could be derived by similar searches targeting a charged Higgs decaying into $tb$
but produced via a different channel.
Specifically, in Ref.~\cite{tbtb_CMS} one signal category is focused on a resonantly produced (through light quarks)
charged Higgs boson with the subsequent decay into $tb$.
However, this signature has only been analysed for charged Higgs masses $\gtrsim 800\GeV$ due to the high
QCD-jet background.
In order to reduce the large background and extend the search down to lower masses, one could require an additional hard $b$-jet in the final state.
This results effectively in a three $b$-tagged jets plus lepton signature.
One of the signal category of \cite{tbtb_ATLAS} (targeting $tb$ production and $tb$ decay) actually requires three $b$-tagged jets and one lepton, but also additional hard jets, 
and the resulting signal efficiency for our production mode is too small to give any relevant constraint.

We now proceed in describing the proposed search for the $qb \to H^\pm \to tb$ topology. The basic selection
requires events with
 at least three $b$-tagged jets and one lepton ($e$ or $\mu$) with transverse momenta of $p_T^b>20\GeV$ and $p_T^l>30\GeV$, respectively. 
 Moreover, events are
required to have missing energy of $E_T^\mathrm{miss}>35\GeV$. The angular separation $\Delta R$ between any two $b$-jets and between any $b$-jet
and the lepton should satisfy $\Delta R>0.4$.
Furthermore, a pseudo-rapidity cut of $|\eta|<2.5$ is applied to the lepton and all $b$-jets.
Finally, the sum $H_T$  of the transverse momenta
of the three $b$-jets and the lepton momentum should satisfy $H_T>350 \GeV$. The dominant background is $t\overline{t}$ production, which we have simulated in \texttt{MadGraph5\_aMC@NLO}
with up to two additional jets in the five-flavour scheme. In the same way we generate the signal $pp\rightarrow H^\pm (\rightarrow tb) + js$
with up to two additional jets in the five-flavour scheme using the 2HDM model file provided in Ref.~\cite{Degrande:2014vpa}. 

The resulting Les Houches Event (LHE) file is fed into
a \texttt{PYTHIA 8.2} standalone version.
Jets are matched using the MLM jet matching algorithm~\cite{Hoeche:2005vzu,Mangano:2006rw,Alwall:2011uj}. The jet finding is performed with \texttt{FastJet}~\cite{Cacciari:2005hq,Cacciari:2011ma} using the anti-$k_T$ algorithm with
a radius parameter of $R=0.6$. The $K$ factor is determined by dividing the NNLO cross section, see Ref.~\cite{Czakon:2011xx}
and references therein, by the leading order cross section after jet merging. This yields $K\approx 1.6$. We implement jet tagging as follows: the closest jet (within a given radius $\Delta R$) to a $b$-quark that originates from the initial hard process is marked as a $b$-jet, and similarly the closest jet to a $c$-quark that originates from initial hard process is marked as a $c$-jet.  Every other jet are potential mistag candidates, with an assigned mistagging probability taken from the Delphes detector cards that use the operating point of Ref.~\cite{ATLAS:2015dex} with 70\% $b$-tagging probability. Similarly $b$- and $c$-jets are identified as such by multiplying with a tagging probality. We have validated our \texttt{PYTHIA} code (in particular the jet tagging algorithms) by comparing to the simulations obtained for similar signatures by ATLAS~\cite{ATLAS:2015nkq} and CMS~\cite{CMS:2012jea}, for which we obtain excellent agreement (see Appendix~\ref{validation2} for details). 

Our results for the LO partonic cross sections are given in Tab.~\ref{comparison}
for a benchmark point with 
$g_{tb} = 0.6, g_{cb} = 0.4, m_{H^\pm} = 300 \GeV$. This choice corresponds to the benchmark point ``BP1'' in Ref.~\cite{Ghosh:2019exx}, and we essentially agree on the values for signal and  dominant background ($t \overline{t}$ + jets), although we find small discrepancies for sub-leading backgrounds.

\begin{figure}[H]
\centering
    \includegraphics[width=0.49\textwidth]{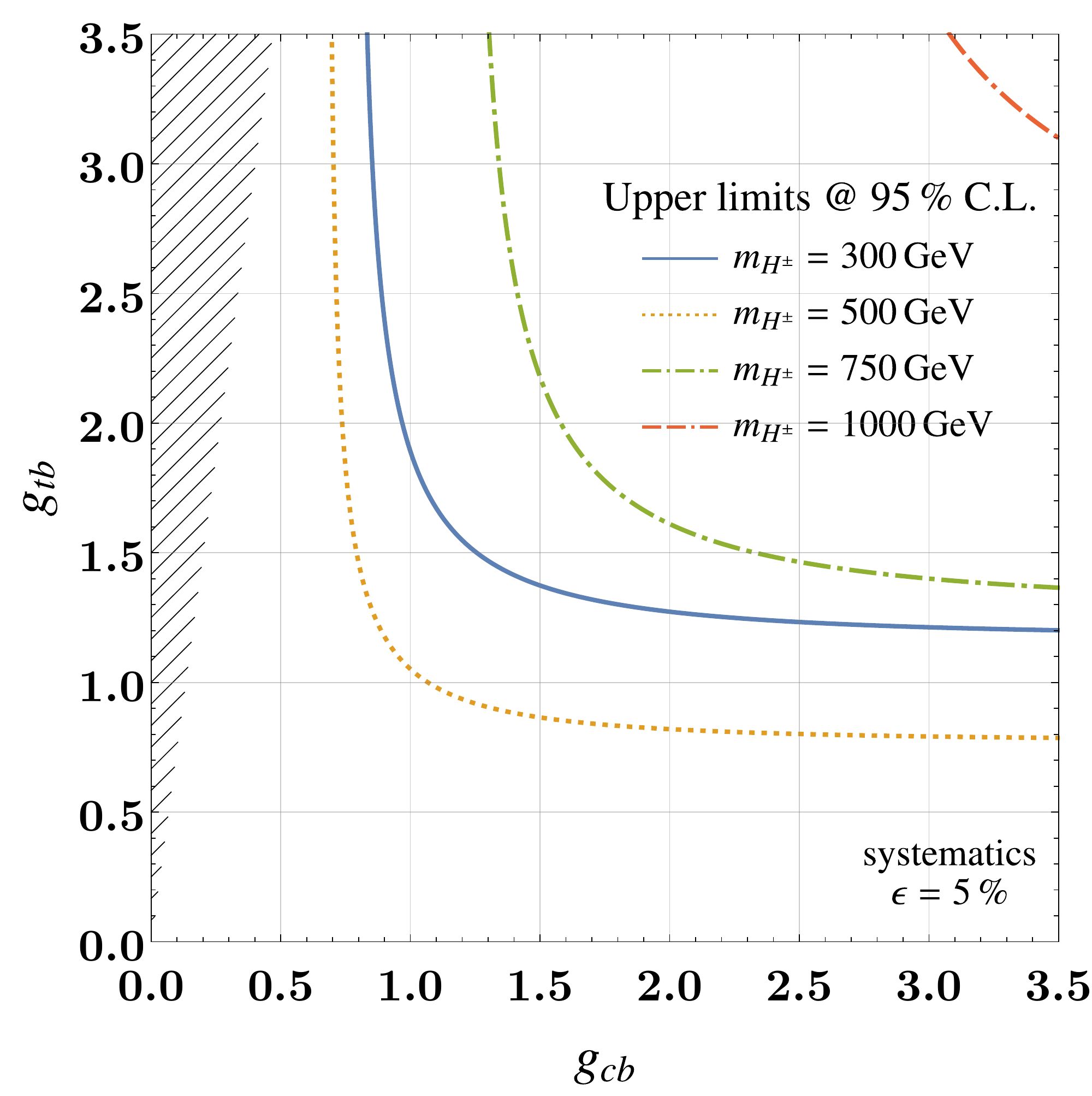}
    \includegraphics[width=0.49\textwidth]{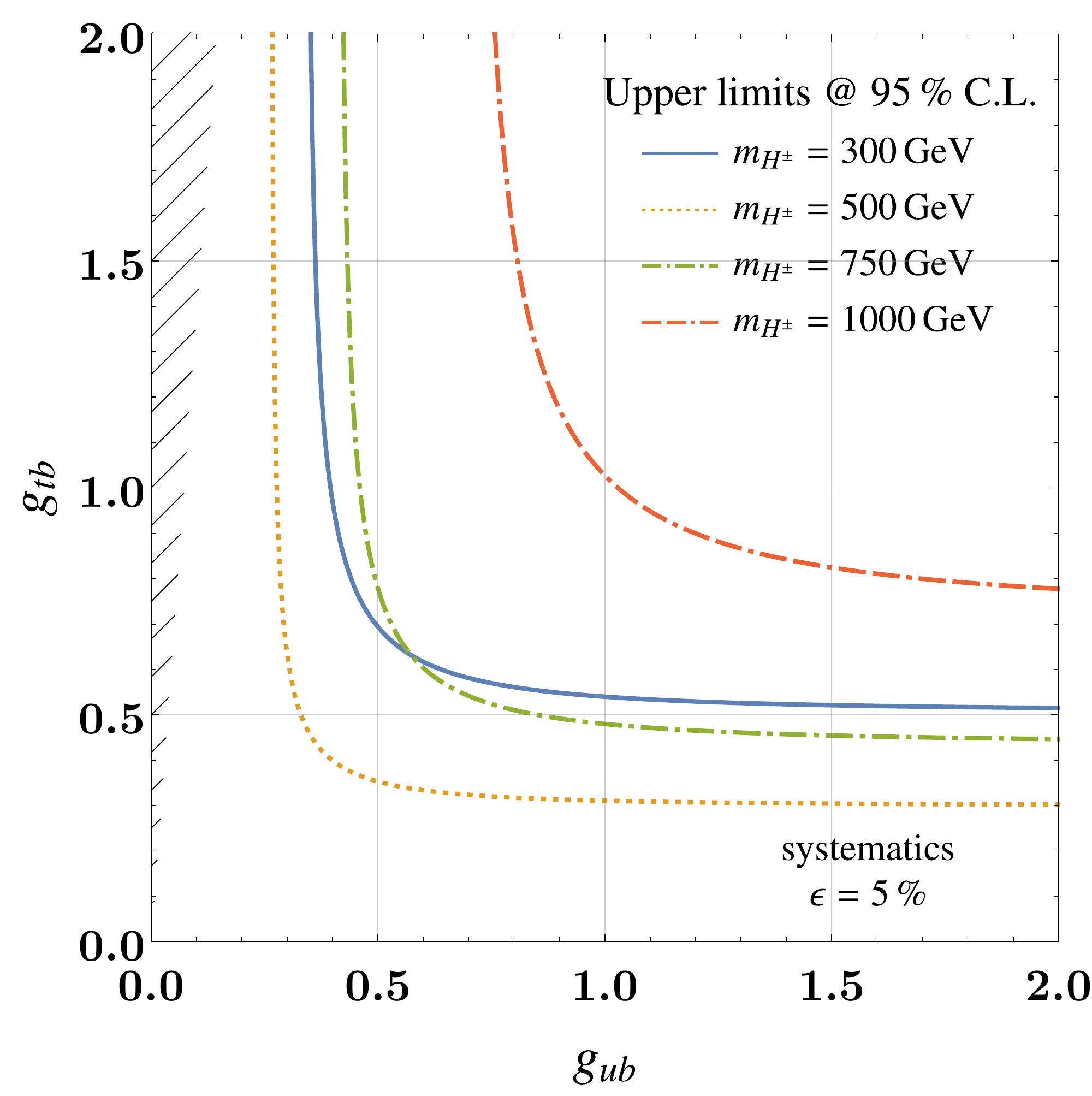}
    \caption{Expected $95\%$ CL upper limits  in the plane of charged Higgs couplings $g_{tb}$--$g_{cb}$ (left panel) and $g_{tb}$--$g_{ub}$ (right panel) 
        for different charged-Higgs masses and $\sqrt{s} = 14\TeV \,@\,139\,\mathrm{fb}^{-1}$. When the ratio of $g_{tb}/g_{cb}$ ($g_{tb}/g_{ub}$) becomes so large that charged Higgs production is no longer dominated by $ug$ ($cg$) fusion, the  bounds from the proposed analysis loose their validity. We indicate this region, roughly defined when associated production with top quarks makes up more than $10\%$ of total charged Higgs production, with a hatching.}
    \label{fig:qb tb}
\end{figure}

\begin{table}[h]
\center
    \begin{tabularx}{1\textwidth}{ccccccc}
    \toprule
                           & $t\overline{t} + 2j$  & $Wt + 2j$             & $tj + 1j$ 
                           & $t\overline{t}h$ & $t\overline{t}Z$ & $2j + H^{\pm}\rightarrow tb$ \\
    \midrule
   $\sigma_{\rm had}^\mathrm{inclusive}$ & $614\,$pb         & $72\,$pb    & $218\,$pb
                           & $480\,$fb         & $709\,$fb   & $28\,$pb  \\
   \\
    $n_{\rm events}$       & 34567910       &  37350 & 383497
                           & 10000     & 10000 & 133247 \\
        $\geq 3j$, $p_T^{b}$, $\eta_b$
                           & 33535695        &  34901  & 301202
                           & 9973     & 9925 & 116646 \\
        $\geq 3 b$         & 1425532         &   815  & 4040
                           & 2849    &  1097   & 10946 \\
        $\Delta R_{bb}$    & 1425532         &   815  & 4040
                           &  2849   &  1097  & 10946 \\
        $\eta_l$           & 1258837         &  708   & 3311
                           &  2722   &  1030   & 9496 \\
        $p^{l,\mathrm{veto}}_T$
                           & 776939         &    427  & 1493
                           &  2026    & 745 & 5278 \\
        $\Delta R_{lb}$,
        $p_T^l$            & 228213          &   131   & 180
                           &  675    &  257   & 1036 \\
        exactly $1$ lepton & 210654          &   119   & 176
                           &  534     &  218  & 1020 \\
        $E_T^{\rm miss}$   & 143130          &   74   & 104
                           &  363     &  155 & 630 \\
        $H_T$              & 45184          &    25   & 23
                           &  181      &  76  & 71 \\
    \midrule
 $\sigma_{\rm had, LO}$ [fb]   & $802\pm 4$     & $49\pm 10$      & $13\pm3$ 
                                       & $8.7\pm 0.6$    & $5.4\pm0.6 $   & $14\pm 2$ \\
    \bottomrule
    \end{tabularx}
\caption{Cutflow of leading order partonic cross sections after merging in the 5FS. \label{comparison}}
\end{table}
In order to study the experimental sensitivity of the proposed analysis, we have used our \texttt{PYTHIA} code to perform a ``cut and count'' analysis, and calculate the statistical significance $Z$
according to
$
    Z \approx S/{\sqrt{S + B + (\epsilon  B)^2}}
    \,,
$
with the number of signal events $S$, background events $B$ and relative systematic background uncertainty $\epsilon$. This analysis is performed  for each charged-Higgs mass point, calculating the number
of signal events as a function of the couplings $g_{tb}$ and $g_{cb}$ by a simple rescaling. In this way one can calculate the expected\footnote{That is the limits one could obtain if only background was observed.} $95\%$ CL exclusion limits in the plane of couplings for a given Higgs mass, and we show the exclusion curves  for $m_{H^\pm} = 300, 500, 750, 1000 \GeV$ in the left panel of Fig.~\ref{fig:qb tb}.

\begin{figure}[H]
\centering
    \includegraphics[width=0.49\textwidth]{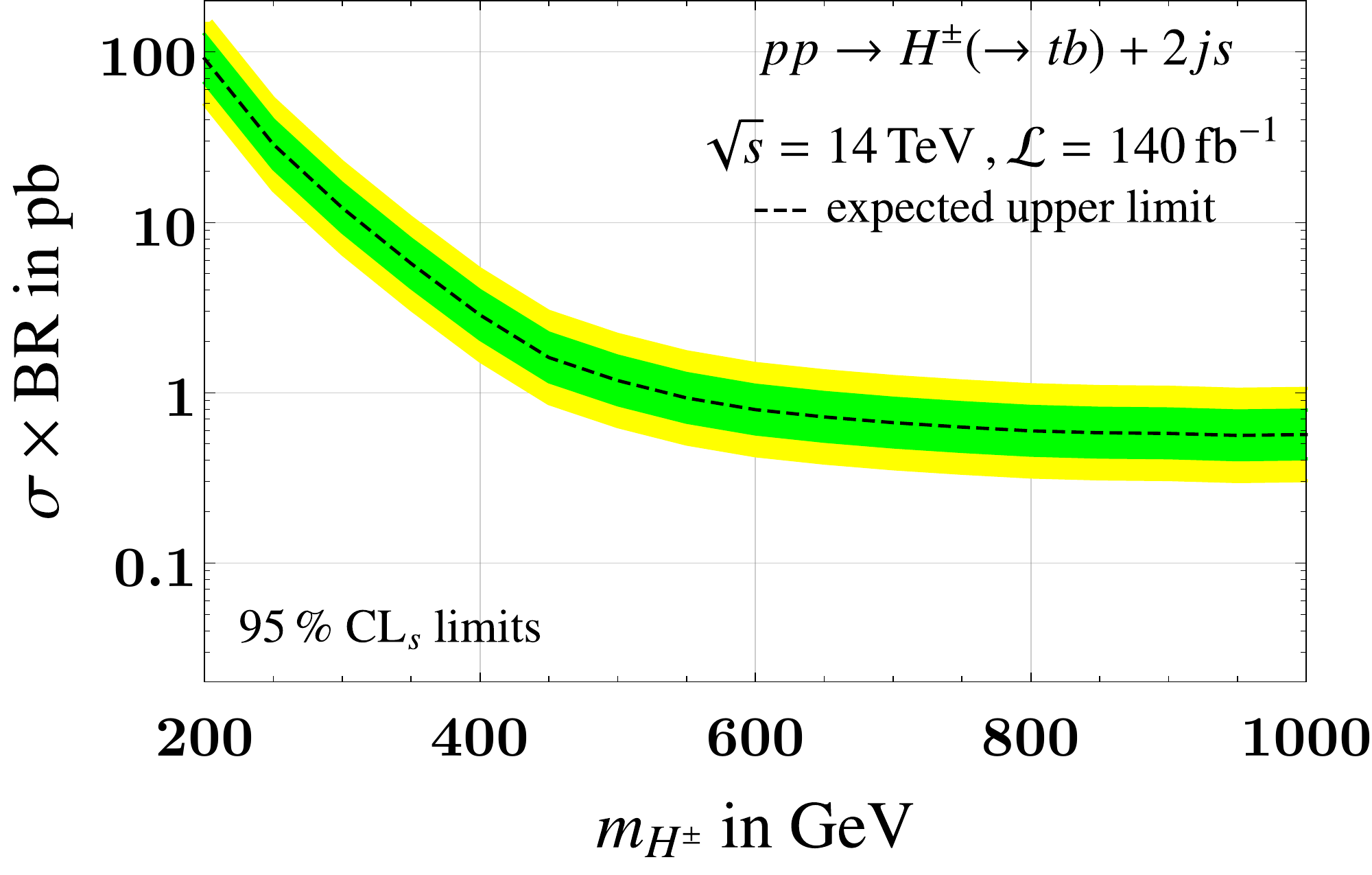}
      \includegraphics[width=0.49\textwidth]{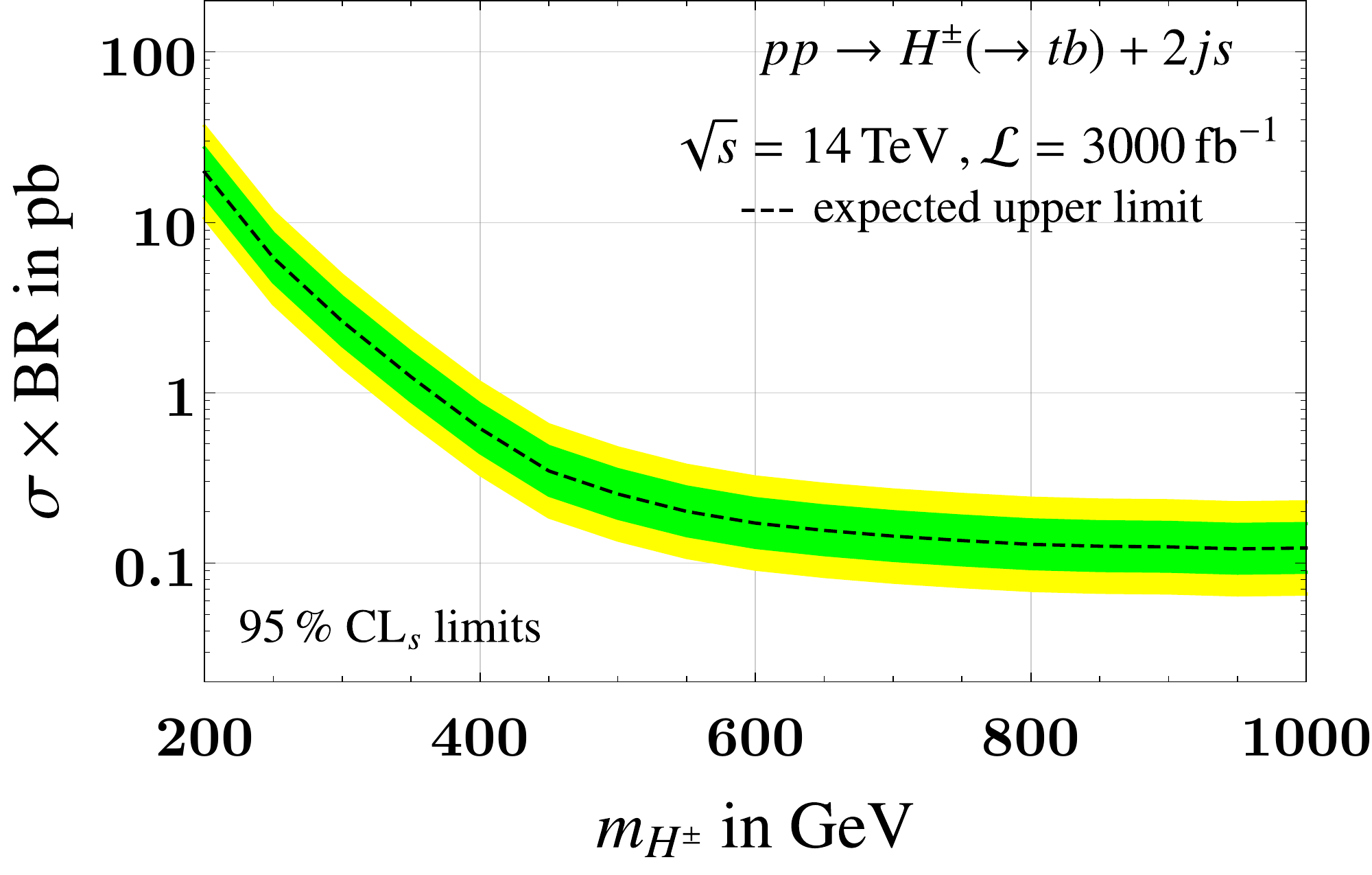}
    \caption{Expected upper limits at $95\%$ CL$_s$ on the production cross section times branching ratio
    $\sigma(gq \rightarrow b H^\pm) \times \mathrm{BR}(H^\pm \rightarrow tb)$  as a function of the charged Higgs mass. Shown in green (yellow) are the projected limits corresponding to background fluctuations of 68\% (95\%). }
    \label{fig:qb tb sigma X BR}
\end{figure}

In the same way we can calculate exclusion curves on the product $g_{tb} g_{ub}$, shown in the right panel of Fig.~\ref{fig:qb tb}. These limits are stronger simply because of the larger parton luminosities for up-quarks as compared to charm quarks, which control charged Higgs production through $ug$ ($cg$) fusion. Note that for large $g_{tb}$ the contribution to charged Higgs production from associated production with top
quarks is not negligible anymore, and the constraints from the proposed analysis no longer apply.  We therefore indicate with a hatching the region where associated production with top quarks makes up more than 
$10\%$ in both panels of Fig.~\ref{fig:qb tb}.
We also provide the constraints on the product of production cross section and branching ratio in Fig.~\ref{fig:qb tb sigma X BR}. The dashed line indicates the 95\% ${\rm CL}_s$ limits that can be obtained if only background was observed, while the green (yellow) bands are the projected  95\% ${\rm CL}_s$ limits corresponding to up- and downward fluctuations of the background at 68\% (95\%).

\begin{figure}[H]
    \centering
    \includegraphics[width=0.495\textwidth]{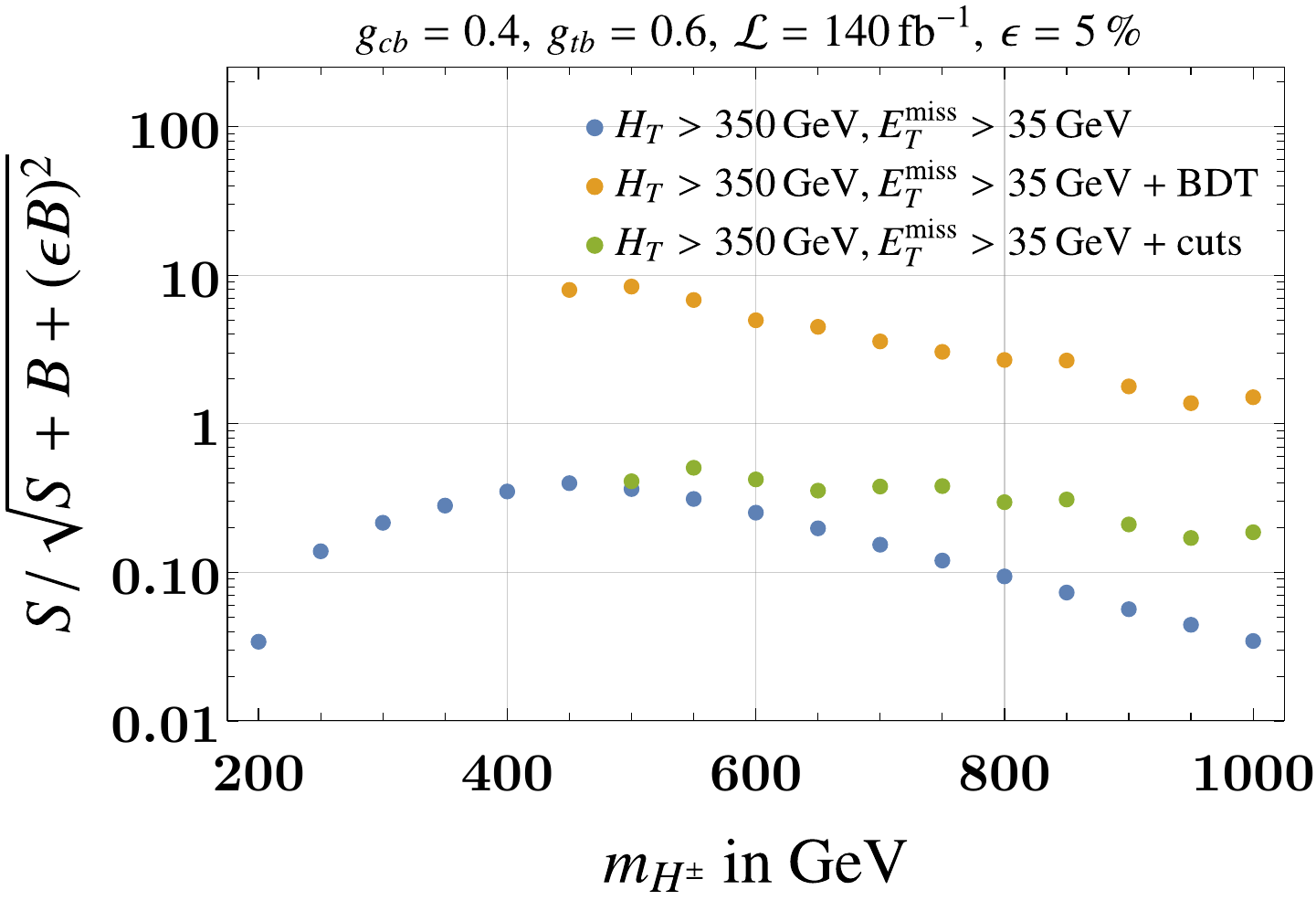}
 \includegraphics[width=0.495\textwidth]{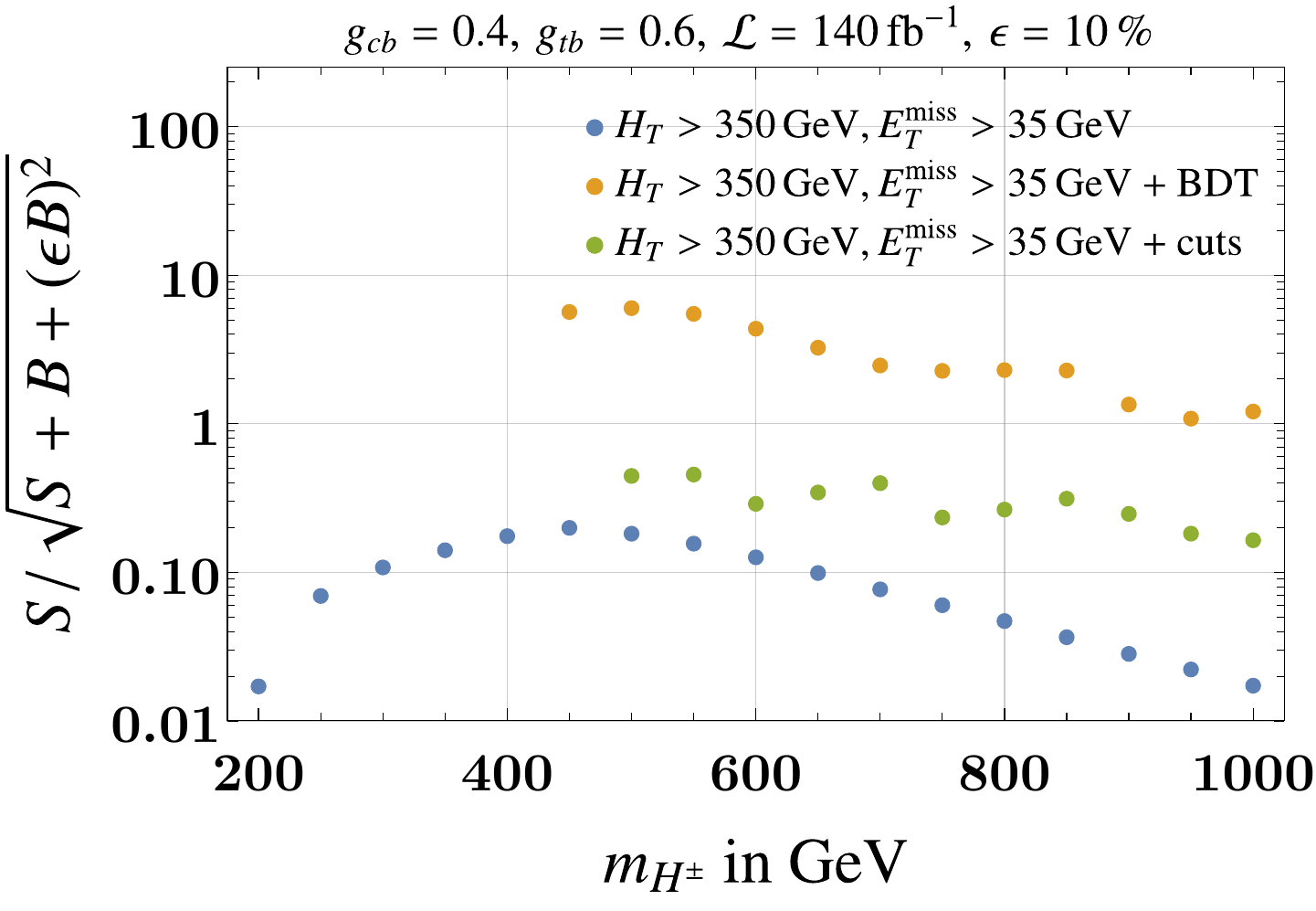}
    \caption{Significance $Z$ as a function of the charged Higgs mass for various cuts. Shown are results for a systematic error of $\epsilon = 5\%$ (left panel) and  $\epsilon = 10\%$ (right panel). Blue points correspond to the cuts of BP1 in Ref.~\cite{Ghosh:2019exx},  green ones to additional cuts on  $p_T, H_T$ and $E_T^\mathrm{miss}$, and orange points correspond to optimised cuts using a boosted-decision-tree algorithm (BDT).}
    \label{fig:optimised cuts}
\end{figure}

Finally, we demonstrate that the sensitivity can be further improved by optimizing the baseline cuts proposed in Ref.~\cite{Ghosh:2019exx}. We take the same basic signal region, i.e.  $\geq 3$ $b$-jets with $p_T\geq 20\GeV$ and exactly 1 lepton with $p_T\geq 30\GeV$, all satisfying $|\eta| < 2.5$, and vary the cuts on missing transverse energy $E_T^{\mathrm{miss}}$ and total transverse momentum $H_T = p_T^l + \sum_{i=1}^3 p_T^{b_i}$. 
\begin{figure}[H]
\includegraphics[width=0.5\textwidth]{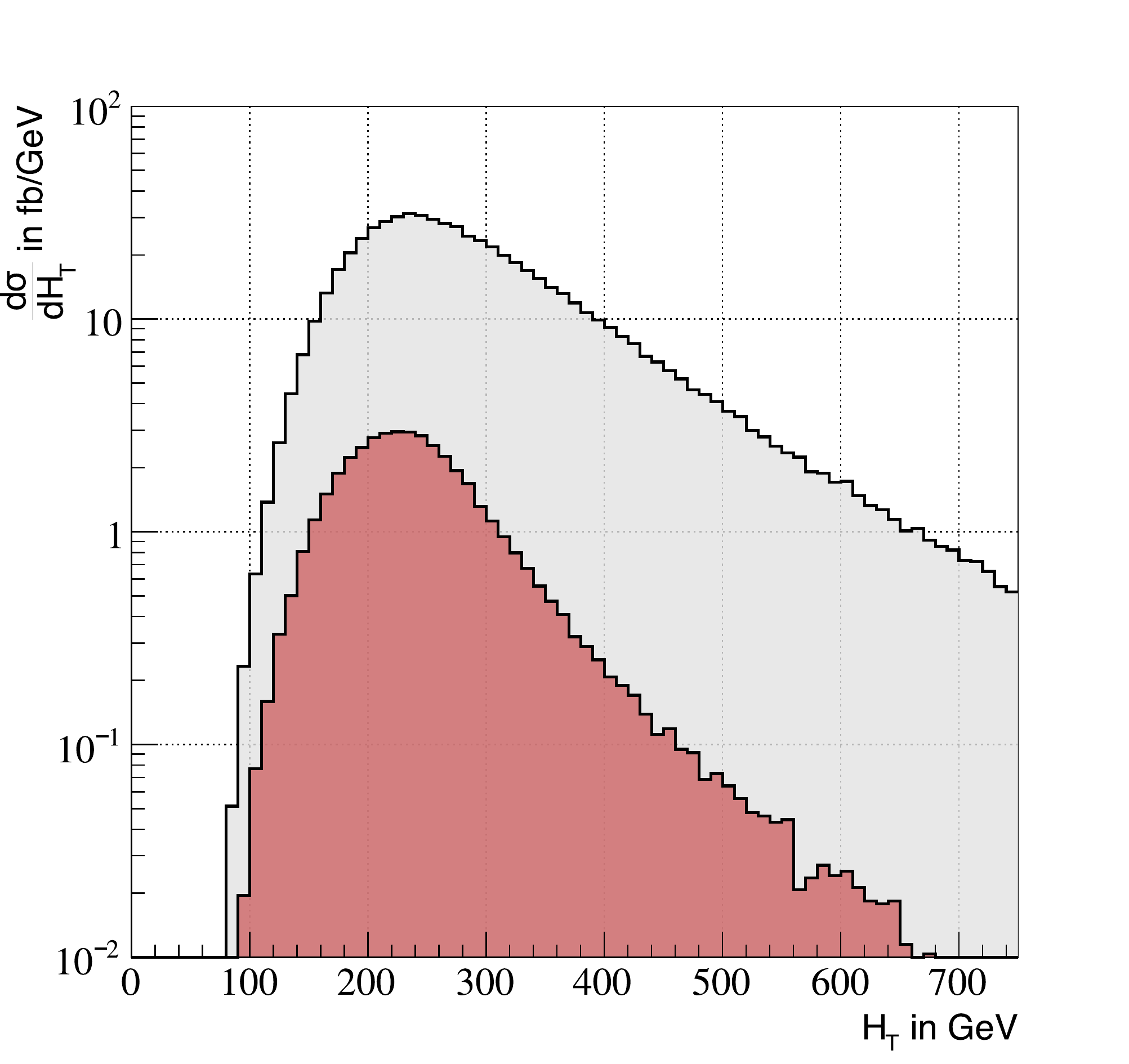}
\includegraphics[width=0.5\textwidth]{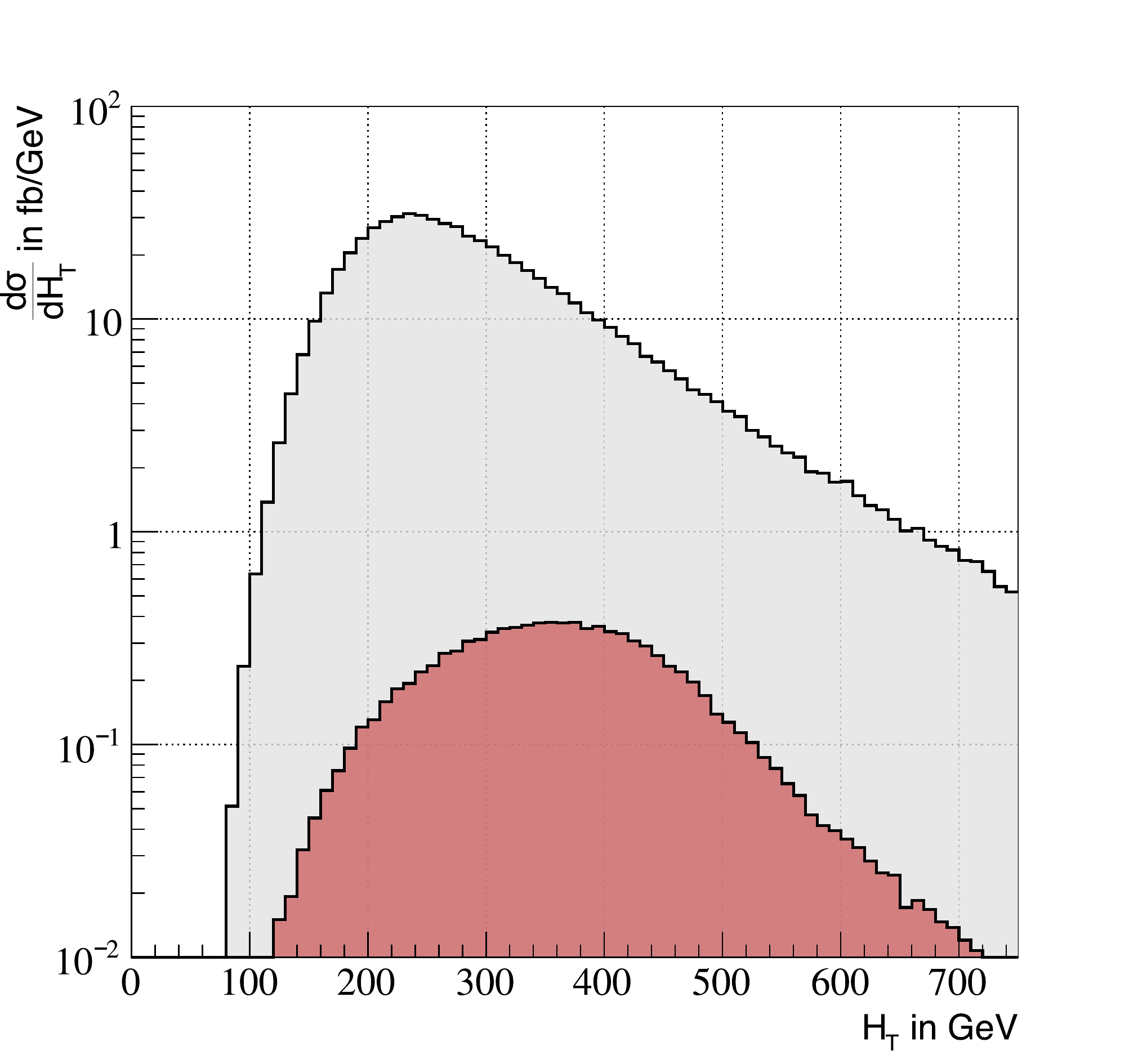}
\caption{Binned differential cross-section in the signal region for $g_{tb} = 0.6, g_{cb} = 0.4$ and $\sqrt{s} =
14 \TeV@140 \, {\rm fb}^{-1}$, as a function of the sum of transverse momenta $H_{\rm T}$ for $m_{H^\pm} = 300 \GeV$ (left panel) $m_{H^\pm} = 500 \GeV$ (right panel).  \label{HistHT}}
\end{figure}
As can be seen from the histograms in Fig.~\ref{HistHT}, a cut on $H_{\rm T}$ can be helpful in distinguishing signal and background for Higgs masses above 500 GeV (we have checked that a cut  on $E_T^{\rm miss}$ has less impact). In the following we consider two procedures to maximize the significance $Z \approx S/\sqrt{S+B + (\epsilon B)^2}$:  we try to strengthen the single cuts on $p_T, H_T$ and $E_T^{\rm miss}$ by hand (``optimized cuts''), and also employ a boosted-decision-tree (BDT) algorithm to find the best cuts (``BDT''), see Appendix \ref{BDT} for more details. We apply these procedures to the couplings of the benchmark scenario BP1 ($g_{tb} = 0.6, g_{cb} = 0.4$), and take as a reference point the basic cuts 
for $m_{H^\pm} = 300 \GeV$, $H_T>350\GeV$ and $E_T^\mathrm{miss}>35\GeV$.  In Fig.~\ref{fig:optimised cuts} we calculate the sensitivity for 140\,fb$^{-1}$ as a function of the charged Higgs mass for this reference point (shown in blue). This is compared to an analysis with additional cuts (in green) and the optimized BDT analysis (in orange) We take into account systematic errors through the parameter $\epsilon$, and show results for the cases $\epsilon = 5\%$ (left panel) and   $\epsilon = 10\%$ (right panel). As one can see from this figure the optimized cuts allow a slight gain of sensitivity for Higgs masses above 500\,GeV, while the BDT can potentially increase the sensitivity by an order of magnitude, even for small Higgs masses. We expect that a realistic analysis gives sensitivities that fall between our analyses with additional cuts and the BDT (i.e. between orange and green points in Fig.~\ref{fig:optimised cuts}), provided that we did not underestimate systematic errors.  

\subsection{Charge asymmetry}
\label{chargeA}
In this subsection we explore the possibility of employing charge asymmetry information to further probe at the LHC a charged Higgs 
with flavour violating couplings in the $qb \to H^{\pm} \to tb$ channel. 
Previous studies on charge asymmetries, focussing on the opportunity to use this variable to detect single top signatures or new physics models, include Refs.~\cite{Bowen:2005xq, Ferrario:2008wm, Craig:2011an, Rajaraman:2011rw, Knapen:2011hu, Ko:2012ud, Kumar:2013jgb}. 

For a specific final state including one reconstructed lepton, a charge asymmetry variable $\mathcal{A}_C$ can be constructed as
\begin{equation}\label{eq:charge asym}
\mathcal{A}_C = \frac{N_+ - N_-}{N_+ + N_-} 
\end{equation}
where $N_+$ ($N_-$) denote the number of events with positive (negative) charged lepton.
As we will briefly review in the following, the advantage of using this variable for setting limits (or improving discovery potential) is that many systematic uncertainties simplify in this ratio.

The uncertainty on the charge asymmetry $\mathcal{A}_C$ is calculated using the usual variance formula
for independent variables. In this case, the total statistical uncertainty $\delta^\mathrm{stat}_{\mathcal{A}_C}$ simplifies to
\begin{align}
    \delta^\mathrm{stat}_{\mathcal{A}_C}
    = 2\sqrt{\frac{N_+ N_-}{\left( N_+ + N_- \right)^3}} \,,
\end{align}
since $N_+$ and $N_-$ are Poisson distributed. Also a total systematic uncertainty $\delta^\mathrm{syst}_{\mathcal{A}_C}$ arises from combining the systematic uncertainties on  all processes $p_i$ that yield a non-vanishing charge asymmetry $\mathcal{A}_C(p_i)$
\begin{align}\label{eq:syst error}
    \delta_{\mathcal{A}_C}^\mathrm{syst}
    \approx \frac{\sqrt{\sum_i \left[ \delta_{\sigma(p_i)}^2 \left( \mathcal{A}_C^2(p_i) + \mathcal{A}_C^2 \right) \right] }}{
       \sum_i \sigma (p_i)
    }  \,,
\end{align}
where $\delta_{\sigma(p_i)}$ denotes the systematic uncertainty on the production cross
section of the process $p_i$. Note that the first term, which arises from the uncertainty on $N_+ - N_-$, typically dominates over the second term that arises from propagating the uncertainty on $N_+ + N_-$. 

In the case of a charged Higgs coupled predominantly to either $u b$ or $cb$ and the corresponding $bbb \ell$ signature, as studied in this section, 
the use of the charge asymmetry can be twofold.
On one hand, it can be employed as a new variable to test the pure $u b$ coupling hypothesis, since for such coupling the 
new physics signal is maximally charge asymmetric, contrary to the SM background.
On the other hand, in the optimistic case of a signal discovery in the $bbb \ell$, it can effectively discriminate between $ub$ and $cb$ production
once restricted to the signal events.

\begin{figure}[t]
\centering
    \includegraphics[width=0.7\textwidth]{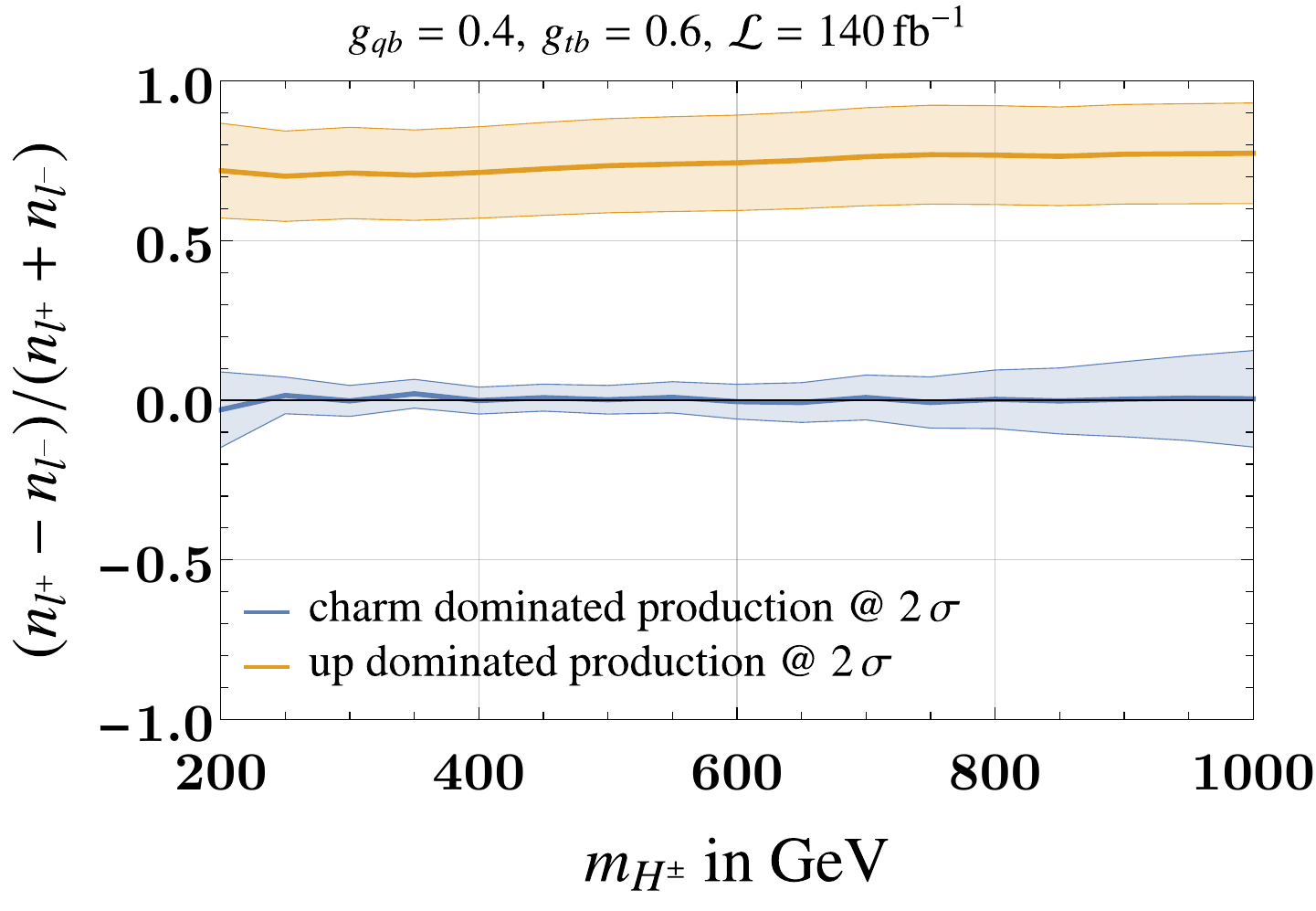}
    \caption{Charge asymmetry for the charged Higgs signature (in $bbb\ell$) for the cases of pure $ub$ (orange) or $cb$ (blue) production, as a function of the charged Higgs mass. The production through the  $ub$ coupling displays a significant charge
    asymmetry due to the initial state valence quark. We also display combined statistical and systematic uncertainties, with the latter assumed to be $10\%$.}
    \label{fig:charge_asy}
\end{figure}

We begin the discussion with the second application of $\mathcal{A}_C$, the signal discrimination.
In Fig.~\ref{fig:charge_asy} we show the charge asymmetry variable for the signal in the case of $ub$ and $cb$ production, for the benchmark scenario BP1 with varying
charged Higgs mass, and we have also included statistical and systematic uncertainties. 
The $ub$ channel is very charge asymmetric because the production mode involves a valence quark.
Note that the value of the coupling is not relevant for the central value of the asymmetry, but the overall cross section will modify the
number of events and hence will impact the statistical error.
We conclude that the $ub$ and $cb$ production can be easily discriminated by using the charge asymmetry variable $\mathcal{A}_C$ for a sufficient number of observed events. 

We proceed by analyzing the constraining potential of the charge asymmetry variable in the case of pure $u b$ coupling.
In this case $\mathcal{A}_C$ can be useful since the main background is $t \bar t$ production, which is mainly charge symmetric. Indeed only the interference between $q \overline{q}$-initiated processes at NLO and LO yields a negative value of $\mathcal{A}_C^{t \overline{t}}$ (that gets smaller for increasing rapidity cut), while the main contribution to the $t \bar t$ cross-section comes from gluon fusion that is charge symmetric to all orders~\cite{Ferrario:2008wm, Kuhn:2011ri}. Contributions to the charge asymmetry  from $qg$ and $\overline{q}g$ fusion are subleading.

In order to estimate the discovery potential of the charge asymmetry variable, we evaluate it for the case of only SM and for SM+charged Higgs signal, 
fixing the $ub$ coupling to a representative value.
In case of the SM, we also take into account the subleading
contributions from single-top production besides the main $t\overline{t}$
background which is the most relevant one after the $bbb \ell$ selection, cf. Tab.~\ref{comparison}.
The single-top backgrounds have to be taken into account
since the uncertainty on their cross-sections can have a sizeable effect on the error estimate
due to $\mathcal{A}^\text{single-top}_C\approx 1/3$, see Eq.~\eqref{eq:syst error}. We estimate the uncertainties by adding statistical uncertainty and a systematic uncertainty of $100\%$ (single-top) and $200\%$ ($t\overline{t}$) on the SM backgrounds.  The latter error (which we actually deem to be conservative~\cite{Ferrario:2008wm, Craig:2011an}) reflects the fact that we are only partially taking into account the NLO contribution to  $\mathcal{A}_C$ from $t \overline{t}$ production, as we are not considering QCD loop corrections that are mainly responsible for the charge asymmetry in the SM.

\begin{figure}[t]
\centering
    \includegraphics[width=0.7\textwidth]{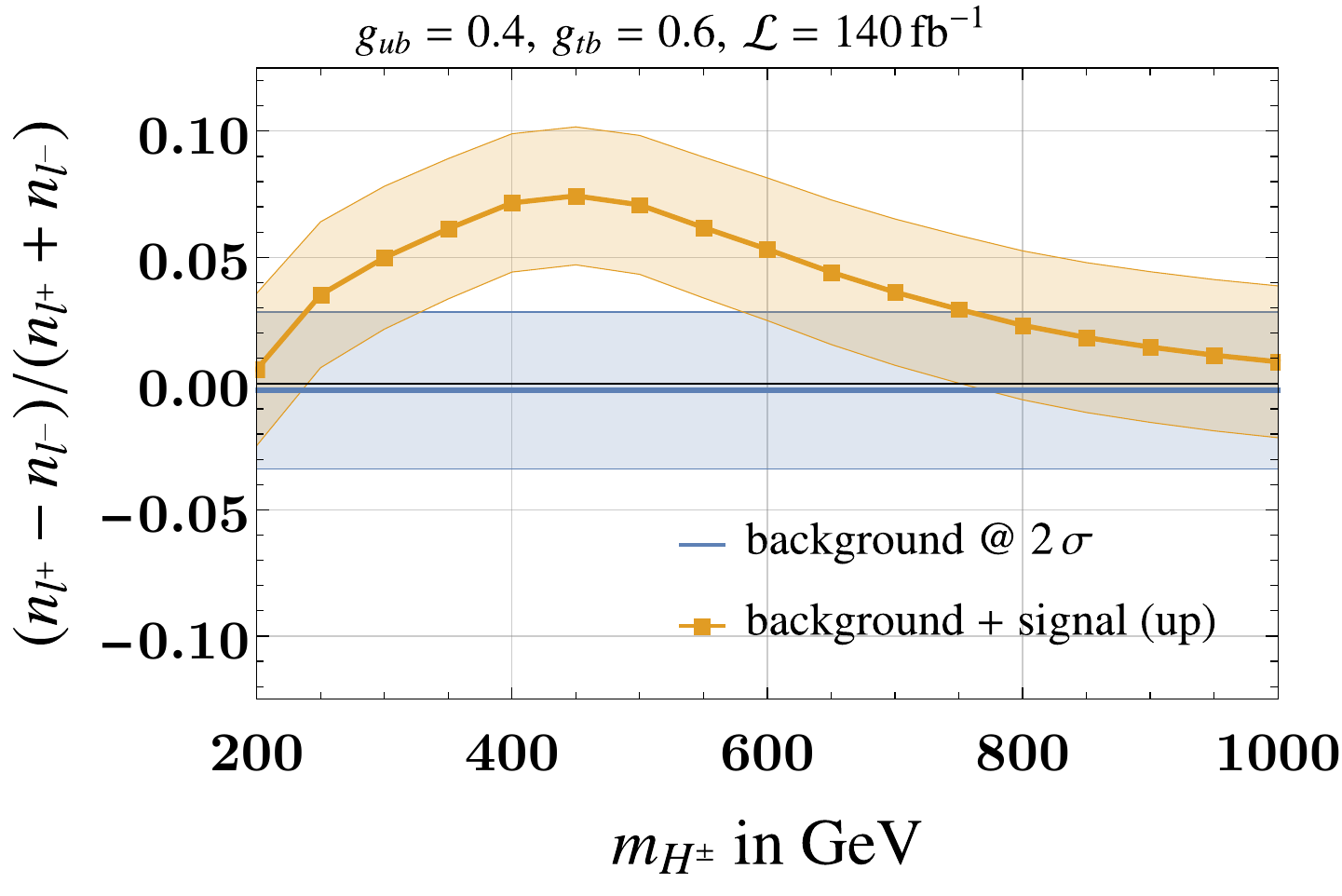}
    \caption{Charge asymmetry comparison between the signal (charged Higgs production through the $ub$ coupling) and the SM background, as a function of the charged Higgs mass,
    for the final state selection $bbb\ell$ explained in the text. 
    We include statistical and systematic uncertainties  in the SM estimate of the asymmetry, see text for details.}
    \label{fig:charge_asy_SM}
\end{figure}

The resulting uncertainty band is shown in Fig.~\ref{fig:charge_asy_SM}.  Within these assumptions, the charge asymmetry variable can for instance 
exclude a charged Higgs
with $ub$ coupling $g_{ub}= 0.4$  in the mass range $325 \GeV \lesssim m_{H^{\pm}} \lesssim 575 \GeV$.
While a detailed quantitative investigation of the systematic uncertainties in the measurement of $\mathcal{A}_C$ is beyond the scope of this paper,
our analysis demonstrates that the use of charge asymmetry variables can provide a promising complementary test of this new physics signature.

\section{Constraints on Charged Higgs Couplings}
\label{Combined constraints}
In this section we summarize the constraints on charged Higgs couplings to $b$-quarks defined in Eq.~\eqref{Lag}, combining the flavor constraints from Section~\ref{Flavor} and the collider constraints for the various production and decay topologies from Section~\ref{Collider}. 

\begin{figure}[h]
\centering
        \includegraphics[width=0.49\textwidth]{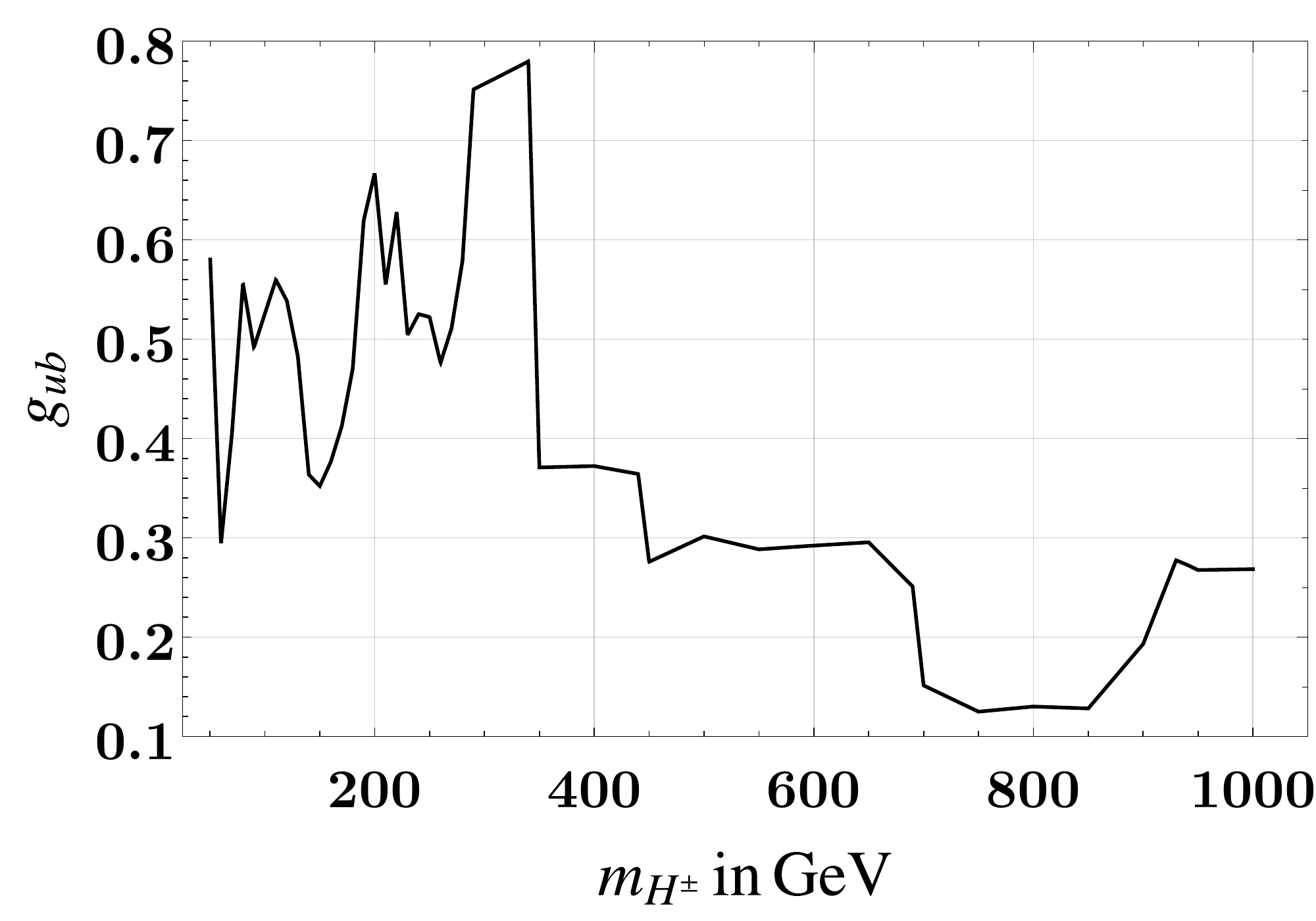}
      \includegraphics[width=0.49\textwidth]{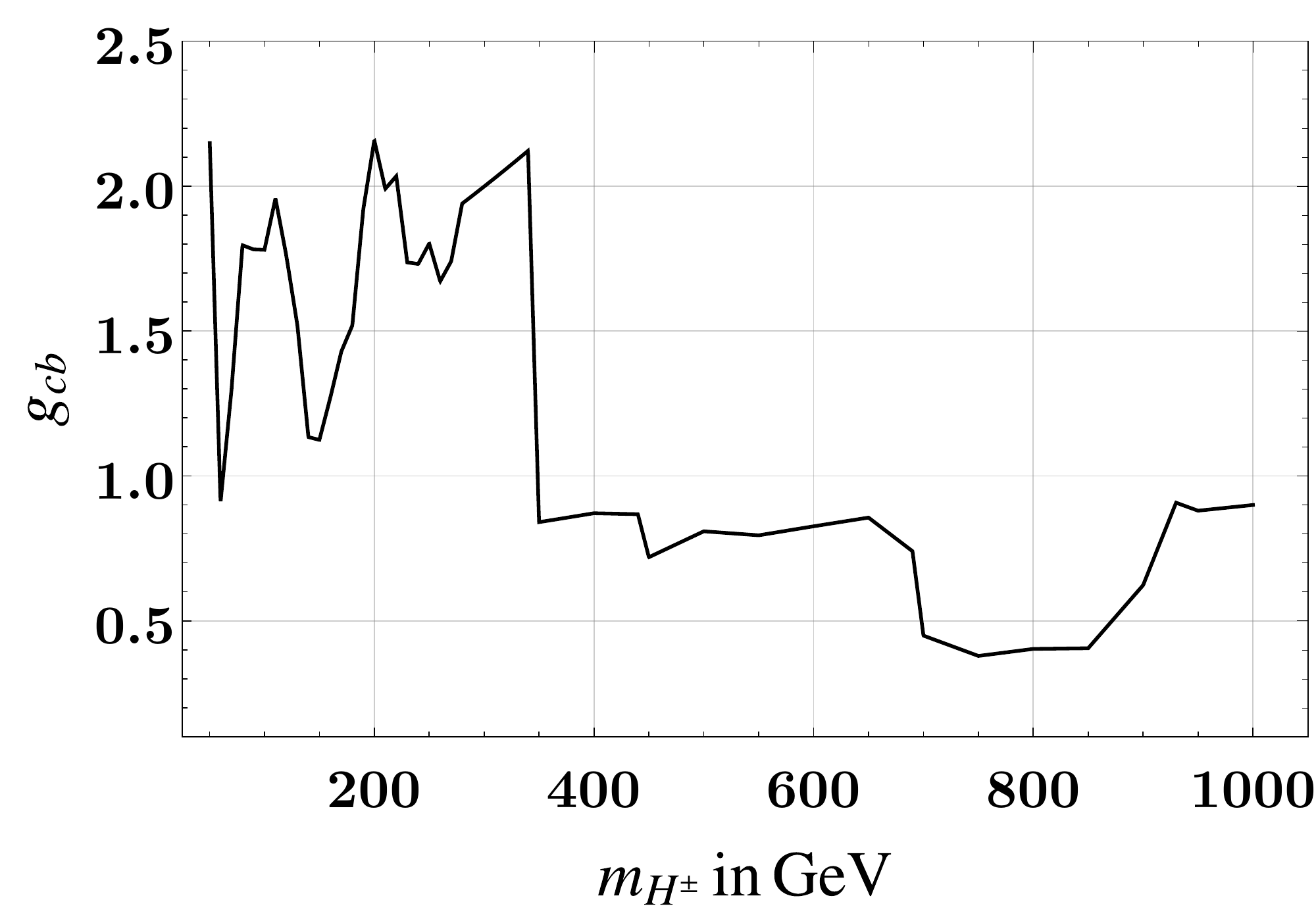}
       \includegraphics[width=0.49\textwidth]{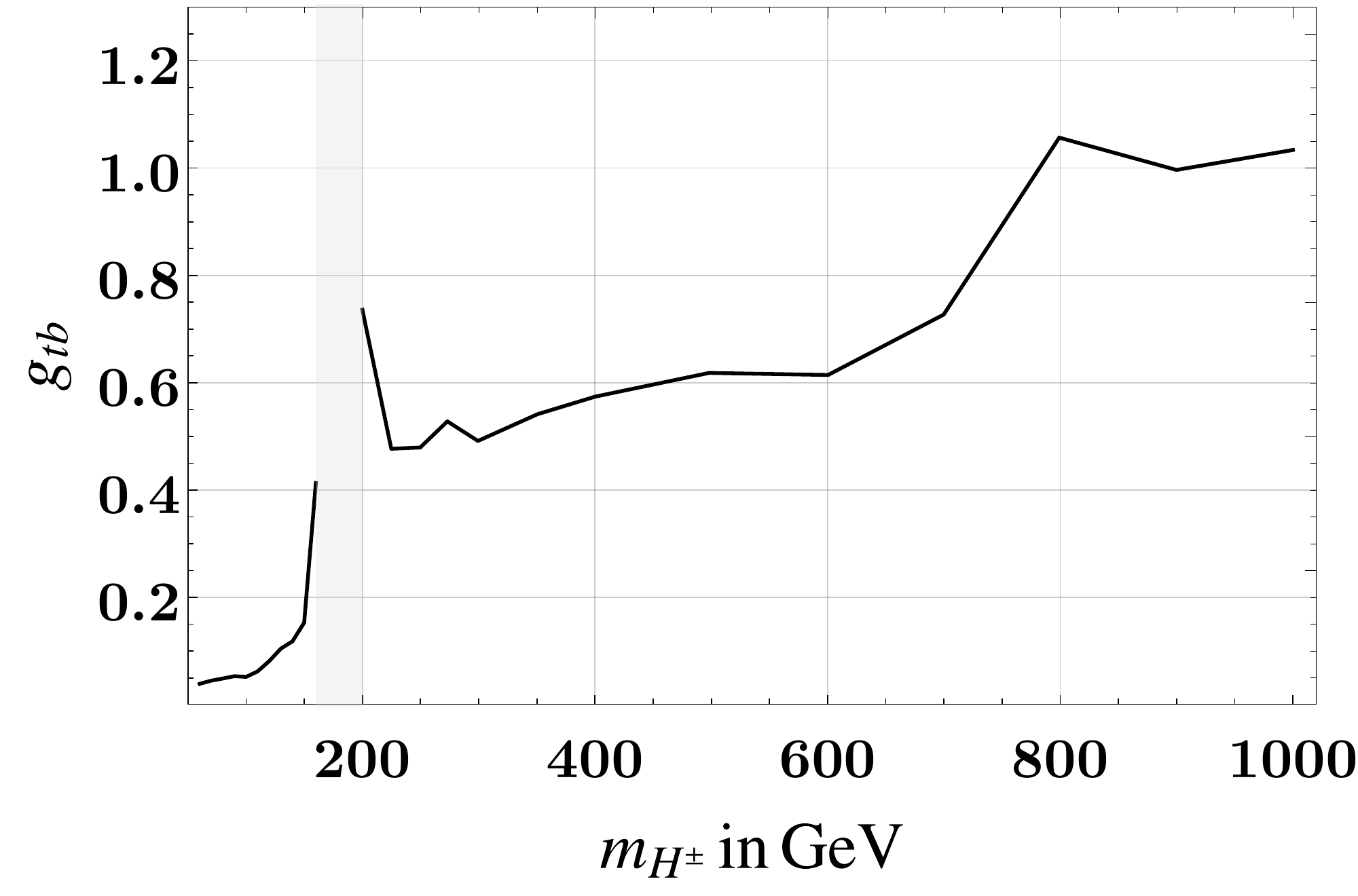}
    \caption{Combined constraints on charged Higgs couplings in the limit where the coupling $g_{ub}$ (left upper panel), $g_{cb}$ (right upper panel) or $g_{tb}$ (lower panel) dominates.}
    \label{fig:1coupling}
\end{figure}

If a single coupling dominates, the constraints are trivially given by Figs.~\ref{fig:tbtb} and \ref{fig:tbcb} for dominant $g_{tb}$ coupling, and Fig.~\ref{fig:dijetsphoton} for dominant $g_{cb}$ or $g_{ub}$ coupling. We summarize these constraints in Fig.~\ref{fig:1coupling}. 
\begin{figure}[h]
\includegraphics[width=0.5\textwidth]{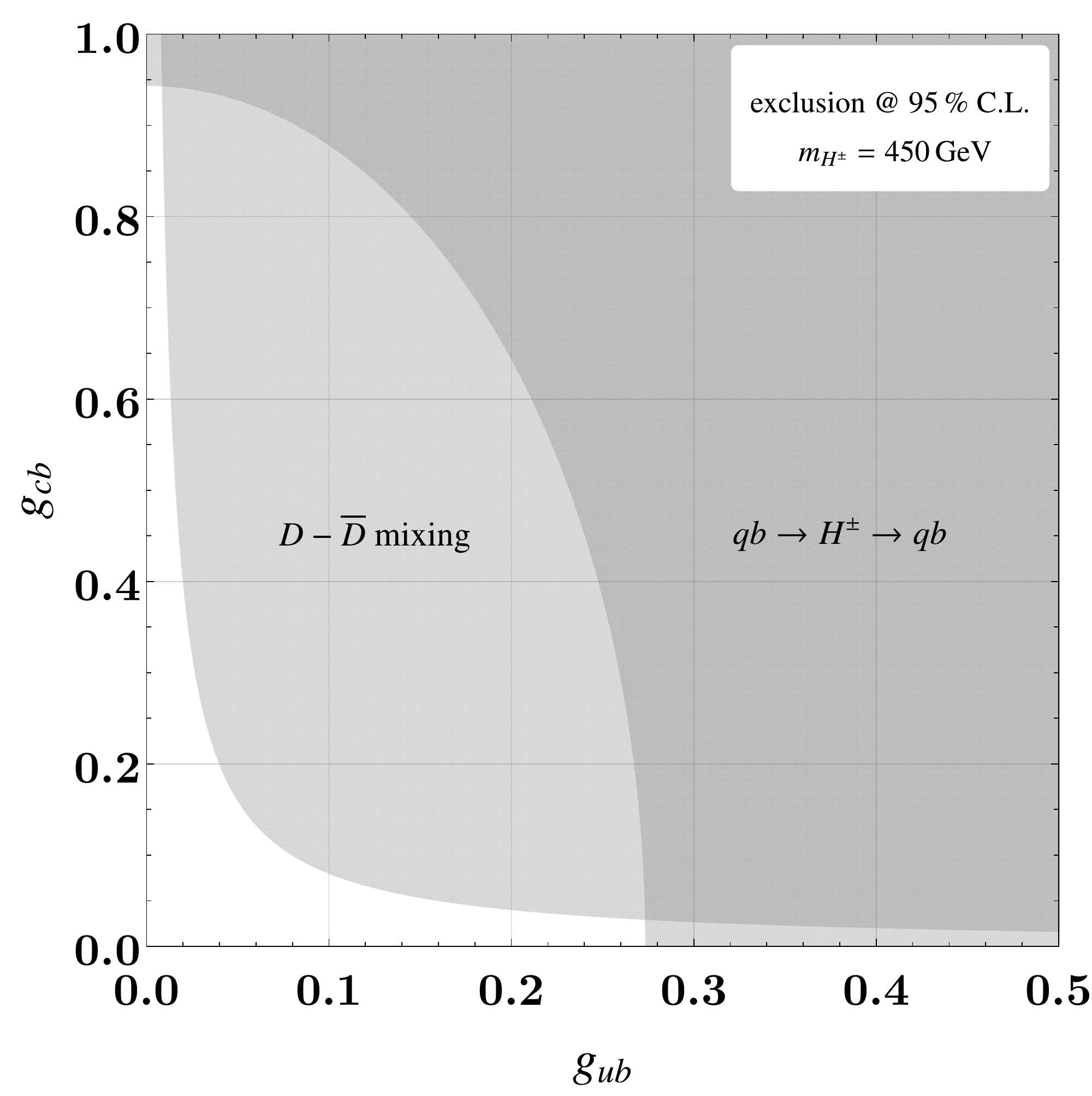}
\includegraphics[width=0.5\textwidth]{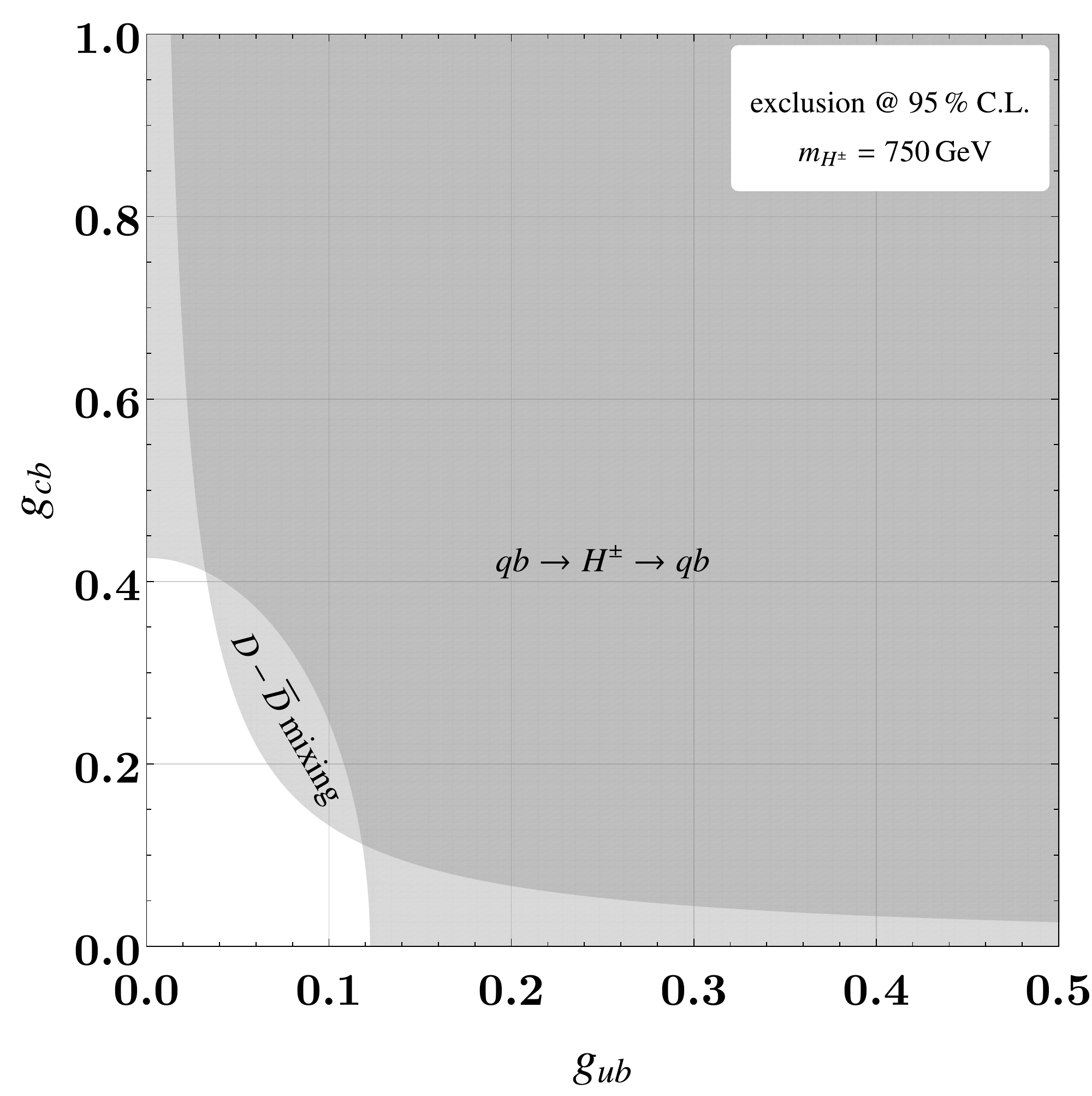}
\caption{Present 95\% CL constraints in the coupling plane $g_{ub}$--$g_{cb}$ for $m_{H^\pm} = 450 \GeV$ (left panel) and $m_{H^\pm} = 750 \GeV$ (right panel). Shown are the constraints from dijet searches~\cite{CMS:2019emo,ATLAS:2018qto,CMS:2018mgb,CMS:2019mcu} denoted by ``$q b \to H^\pm \to qb$'' and flavor constraints denoted by ``$D$--$\overline{D}$ mixing''. 
}
\label{fig:ubcb}
\end{figure}

If both $g_{ub}$ and $g_{cb}$ are sizable, the bounds from flavor physics (cf. Fig.~\ref{fig:D-mixing}) and dijet searches (cf. Fig.~\ref{fig:dijetsphoton}) are complementary , as seen in Fig.~\ref{fig:ubcb}, where these constraints have been overlaid in the $g_{ub}$--$g_{cb}$-plane for two choices of the charged Higgs mass. While for low Higgs masses the flavor constraints dominate for couplings of similar size, these constraints quickly fade for larger Higgs masses, while instead the dijet searches become more effective. Note that in the limit where a single coupling dominates the limits in Fig.~\ref{fig:1coupling} are recovered, but generically the bounds are stronger because the two production channels add up.

\begin{figure}[H]
\includegraphics[width=0.5\textwidth]{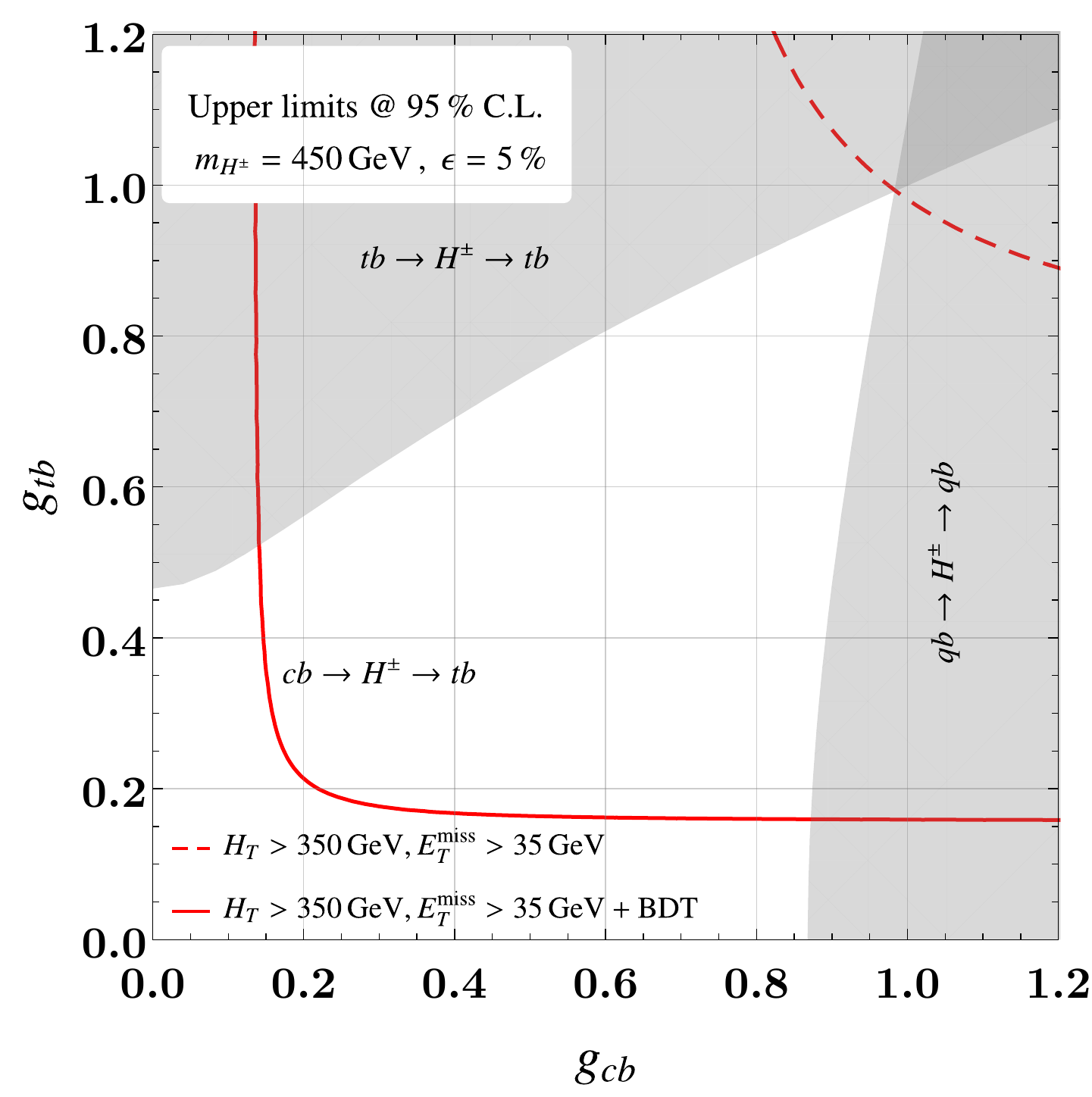}
\includegraphics[width=0.5\textwidth]{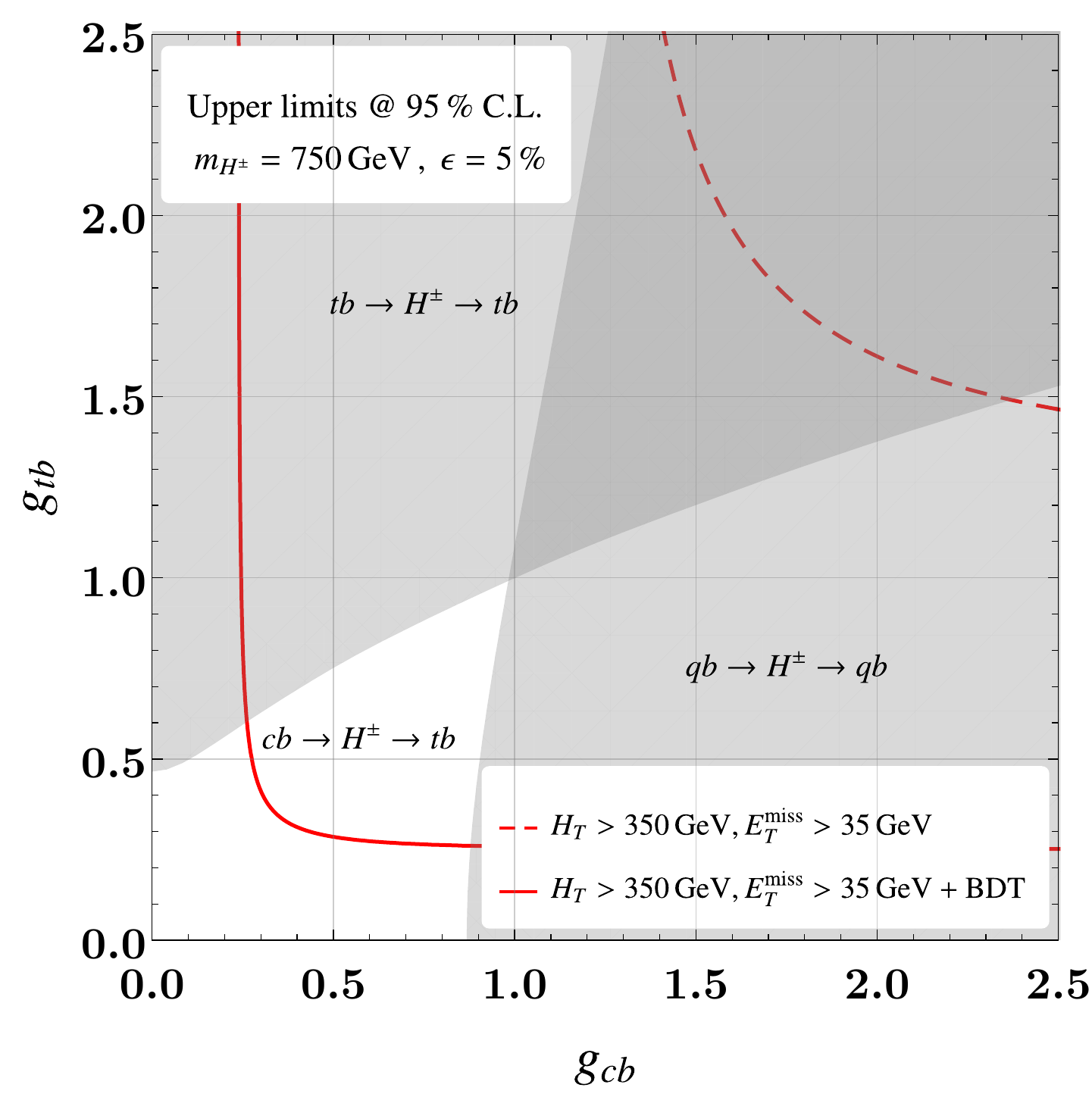}
\caption{\label{cbtb} Expected 95\% CL upper limits in the  coupling plane $g_{cb}$--$g_{tb}$ for $140 \, {\rm fb}^{-1}$ and charged Higgs masses of 450 GeV (left panel) and 750 GeV (right panel). Shown are the  constraints from single coupling searches based on dijet searches~\cite{Aaboud:2019zxd}  denoted by ``$qb \to H^\pm \to cb$'', and searches for top charged Higgs production and decay~\cite{tbtb_ATLAS}  denoted by ``$tb \to H^\pm \to tb$''. The red lines denoted by ``$cb \to H^\pm \to tb$'' indicate the parameter space that could be probed by the search described in Section \ref{qbtb}. 
The dashed red line shows the constraints from applying the indicated cuts, while the solid line shows the constraints obtained from optimized cuts using a BDT algorithm. We assume a systematic uncertainty of 5\%.}
\end{figure}

\begin{figure}[H]
\includegraphics[width=0.5\textwidth]{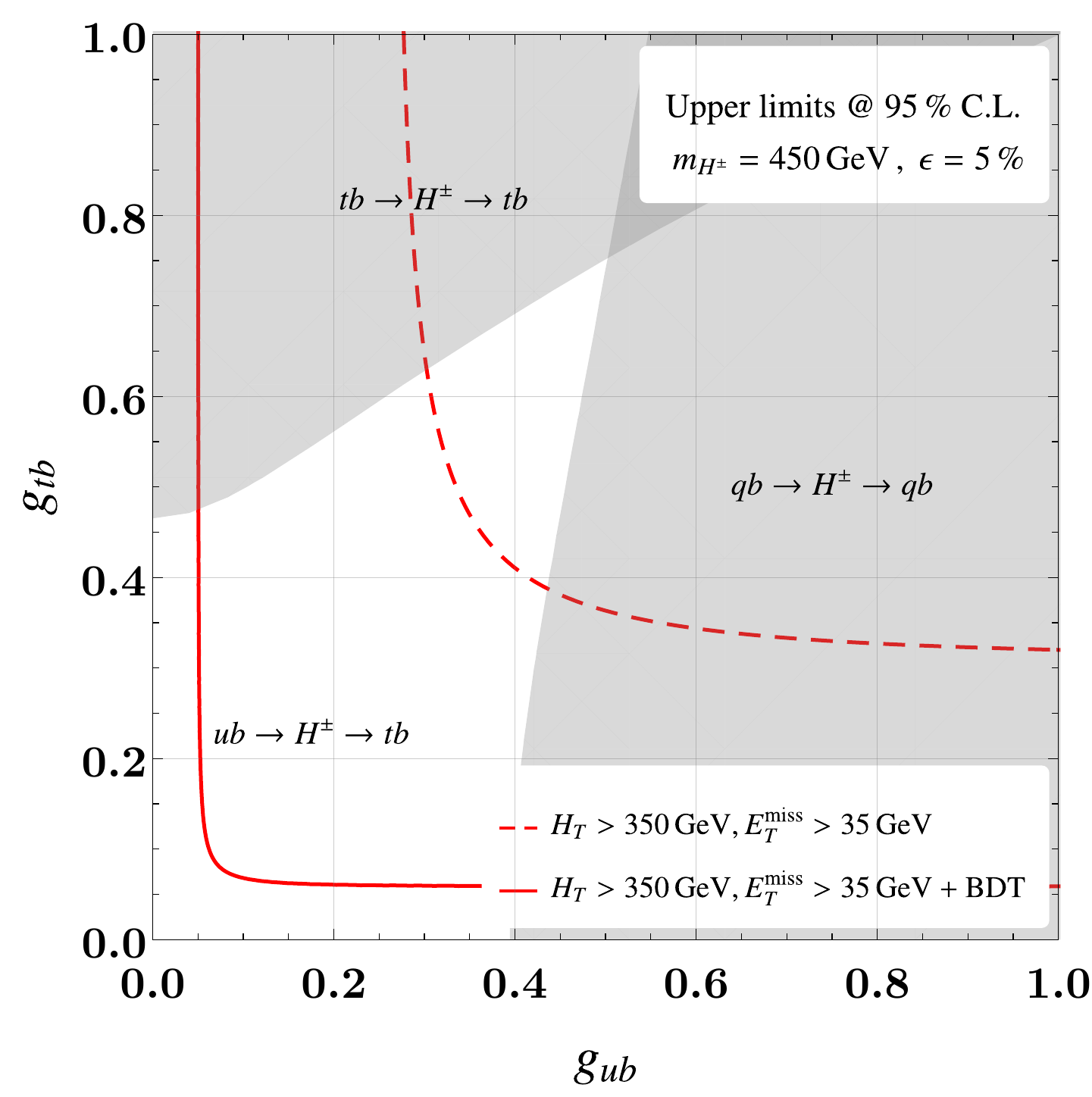}
\includegraphics[width=0.5\textwidth]{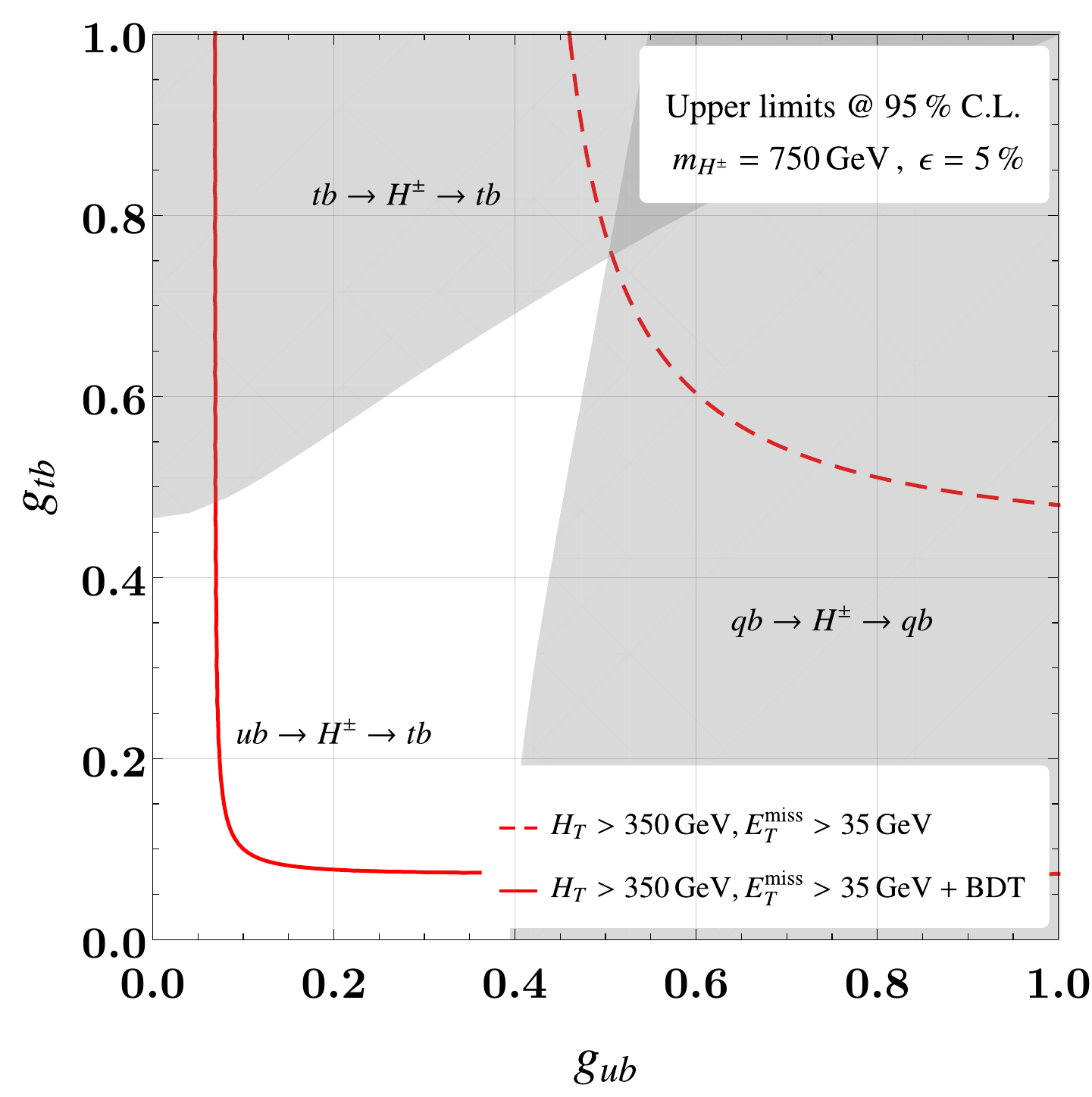}
\caption{\label{ubtb} As Fig.~\ref{cbtb}, but for the coupling $g_{ub}$ instead of $g_{cb}$. }
\end{figure}

At present there is no experimental search that specifically probes the non-trivial interplay of top and light quark couplings. However, as discussed in Section~\ref{qbtb}, strong bounds can be obtained already with  present data  if dedicated searches for the $qb \to H^\pm \to tb$ signature are carried out, as shown in Fig.~\ref{fig:qb tb}. We overlay the expected constraints with the limits on single couplings in Fig.~\ref{cbtb} (Fig.~\ref{ubtb}), which summarizes all relevant bounds in the plane $g_{cb}$--$g_{tb}$ ($g_{ub}$--$g_{tb}$) for $140 \, {\rm fb}^{-1}$ and two choices for the charged Higgs mass. The grey regions indicate the  limits from single coupling searches, either from dijet searches for dominant $qb$ coupling (cf. Fig.~\ref{fig:dijetsphoton}), or the usual searches for top charged Higgs production and decay for dominant $tb$ coupling  (cf. Fig.~\ref{fig:tbtb}). Note that these bounds are loosened when the non-dominant coupling is increased, simply due to the reduction of the relevant branching ratio. 

\begin{figure}[H]
\includegraphics[width=0.5\textwidth]{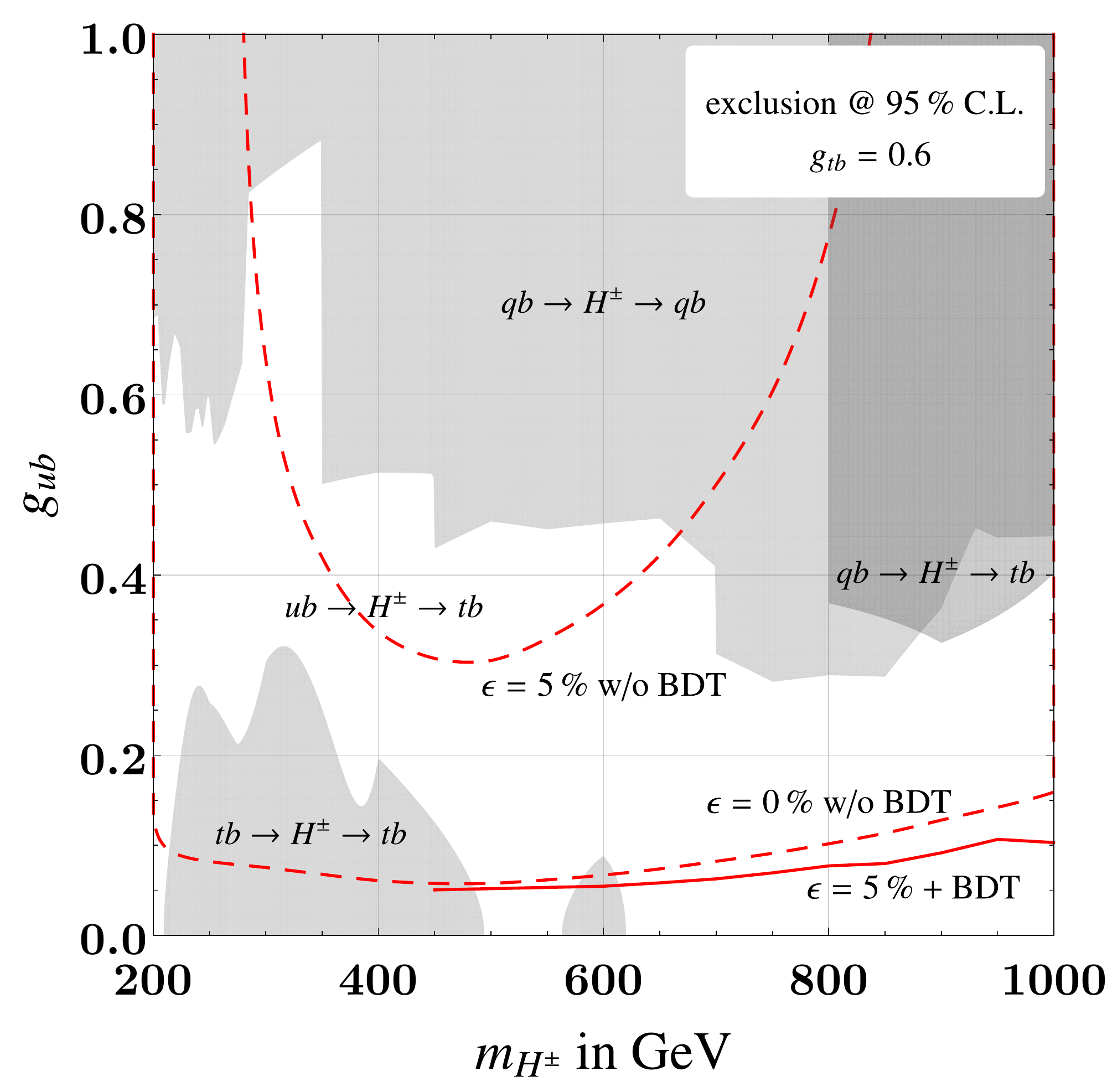}
\includegraphics[width=0.5\textwidth]{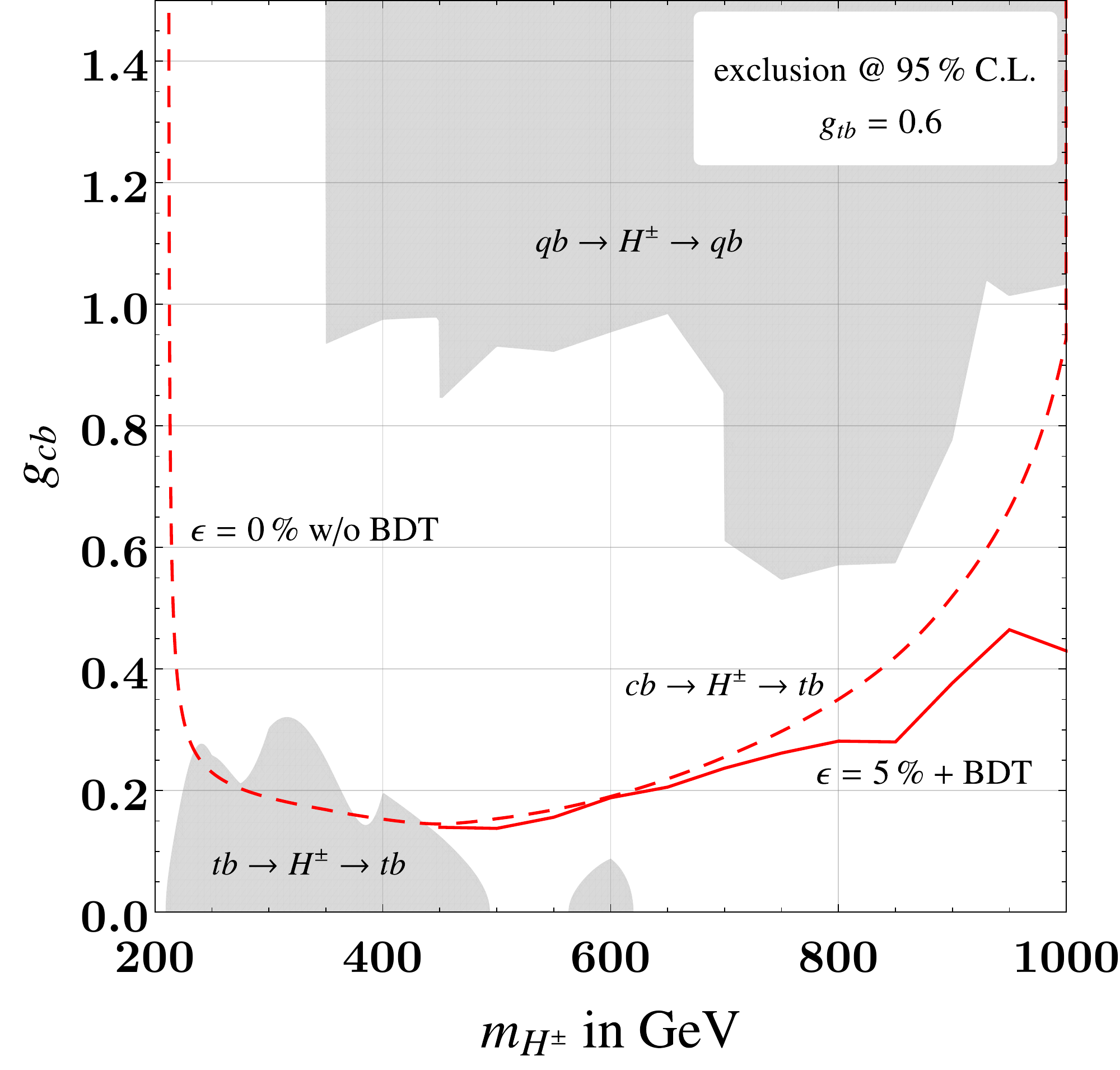}
\caption{\label{qbmH} Expected 95\% CL exclusion limits in the plane $m_{H^\pm}/g_{ub}$ (left panel) and $m_{H^\pm}/g_{cb}$ (right panel) for fixed $g_{tb} = 0.6$. Shown in gray are the  constraints from single coupling searches based on dijet searches~\cite{Aaboud:2019zxd}  denoted by ``$qb \to H^\pm \to qb$'',  searches for top-associated charged Higgs production and decay~\cite{tbtb_ATLAS}  denoted by ``$tb \to H^\pm \to tb$'', and 
the constraints from resonant charged Higgs production decaying into $tb$~\cite{tbtb_CMS} denoted
as ``$qb \to H^\pm \to tb$'' . The latter are  relevant only for $g_{ub}$ and large Higgs masses. The red lines indicate the parameter space that could be probed by the search described in Section \ref{qbtb}, denoting the constraints from applying the cuts in Fig.~\ref{cbtb} with (solid) or without BDT optimization (dashed), for different assumed systematic errors $\epsilon$.}
\end{figure}

The red contours instead indicate the parameter space that could be probed by the dedicated search for $qb \to H^\pm \to tb$  as discussed in detail in Section \ref{qbtb}.
Dashed red lines show the sensitivity for simple cuts on top of $H_T > 350 \GeV, E_T^{\rm miss} > 35 \GeV$, while the solid red line denote the constraints we obtained from cut optimization used a BDT algorithm. We expect that a realistic analysis carried out by the experimental collaborations will yield constraints between these two curves, and thus has the potential to substantially improve present bounds, in particular for light Higgs with couplings to $ub$ and $cb$ quarks of similar size. Therefore such searches will probe a significant portion of previously unchartered parameter space, which can be seen from  Fig.~\ref{qbmH}, where we show the same lines as in Figs.~\ref{cbtb} (Fig.~\ref{ubtb}), but in the plane $m_{H^\pm}$--$g_{cb}$ ($m_{H^\pm}$--$g_{ub}$) for fixed coupling $g_{tb} = 0.6$. This figure also shows that using the BDT is roughly equivalent to reducing the systematic uncertainties by 5\%.

\section{Summary and Conclusions}
\label{Conclusion}
In this work we have systematically studied model-independent constraints on generic charged Higgs couplings to $b$-quarks. Flavor physics ($D$-meson mixing and flavor-violating top decays)  gives only weak constraints, so that LHC searches play a crucial role in probing these couplings. So far ATLAS and CMS have performed dedicated searches only for charged Higgses that dominantly couple to $tb$ quarks, resulting in constraints that we  summarize in Fig.~\ref{fig:1coupling} (lower panel). If instead couplings to $ub$ or $cb$ quarks dominate, one can recast existing dijet searches for leptophobic $Z^\prime$ fields in order to obtain the bounds in the upper panels of Fig.~\ref{fig:1coupling}.  These constraints could be strengthened by new dedicated searches in the following way:
\begin{itemize}
\item Looking for light charged Higgses produced from top decays $t \to b H^\pm (\to ub)$ in a similar way to the existing search for $t \to b H^\pm (\to cb)$,  see Section~\ref{tbqb}. 
\item Extending the dijet searches for $Z^\prime$ vector bosons also with \emph{single} $b$-tagging, in order to improve sensitivity on scenarios with flavor-violating couplings, see Section~\ref{sec:qb qb}. 
\end{itemize}
If couplings to both $ub$ and $cb$ quarks are sizable, there is a non-trivial interplay between LHC searches and flavor physics.  Dijet searches are actually complementary to flavor physics, see Fig.~\ref{fig:ubcb}, as the latter give constraints that quickly decouple with the charged Higgs mass, while the former gain sensitivity for heavy Higgses when hard cuts on transverse jet momenta can significantly reduce SM background. 

Up to now there have been no dedicated searches that look for a charged Higgs that has sizable couplings to both $qb$ and $tb$ quarks. We have discussed in detail how one can profit from their interplay in order to significantly extend the existing reach on the parameter space from single coupling constraints alone,  see Fig.~\ref{cbtb}, Fig.~\ref{ubtb} and Fig.~\ref{qbmH}. Specifically we have proposed two new search strategies:
\begin{itemize}
\item Probing the top decay channels  $qb \to H^\pm \to tb$ using multi-$b$-jet signatures similar to Ref.~\cite{tbtb_ATLAS}, but with lower ($b$-)jet multiplicity, see Section~\ref{qbtb}. 
\item  Employing charge asymmetries that allow to both distinguish $c$- and $u$-production in $qb \to H^\pm \to tb$ and reduce SM background, see Section~\ref{chargeA}. 
\end{itemize}
In summary we think it is worthwhile to develop and carry out dedicated collider searches for all three charged Higgs couplings to $b$-quarks, since the reach of flavor physics is limited and present LHC searches focus on the $tb$ coupling. We suggested various new search strategies that are mainly slight extensions of existing analyses, which hopefully are useful for  experimentalists to design realistic searches that allow to boost the LHC reach on scenarios with generic charged Higgs couplings. 
\section*{Acknowledgements}
We thank Nicola Orlando and Syuhei Iguro for useful discussions. This work is partially
supported by project C3b of the DFG-funded Collaborative Research Center TRR257, ``Particle Physics Phenomenology after the Higgs Discovery''.
AM is supported by the Strategic Research Program High-Energy Physics and the Research Council of the Vrije Universiteit Brussel, and by the ``Excellence of Science - EOS" - be.h project n.30820817.

\appendix

\section{Validation of \texttt{PYTHIA} code used for $qb \to H^\pm \to qb$ analysis}
\label{validation1}
\begin{figure}[H]
    \includegraphics[width=0.49\textwidth]{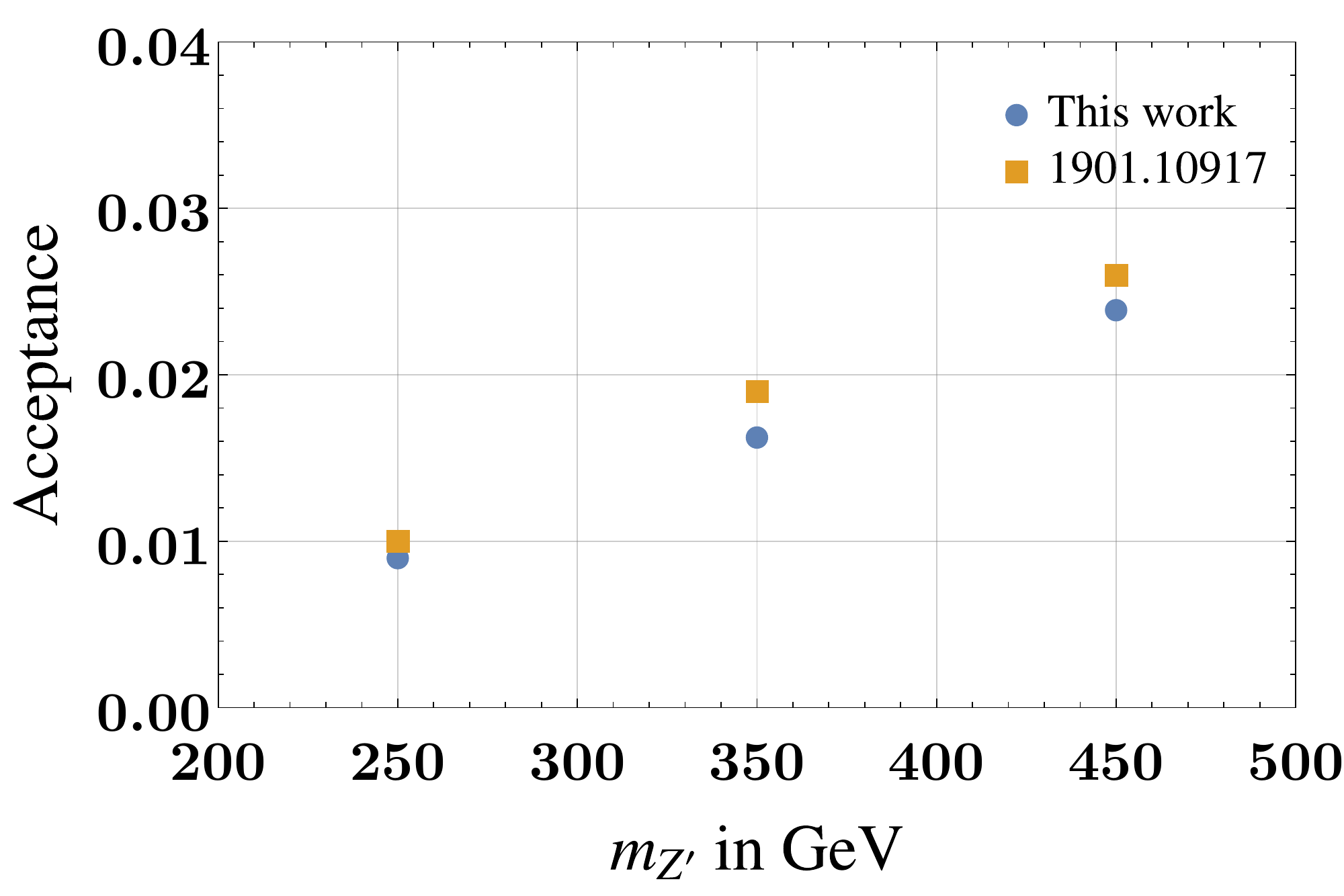}
    \hfill
    \includegraphics[width=0.475\textwidth]{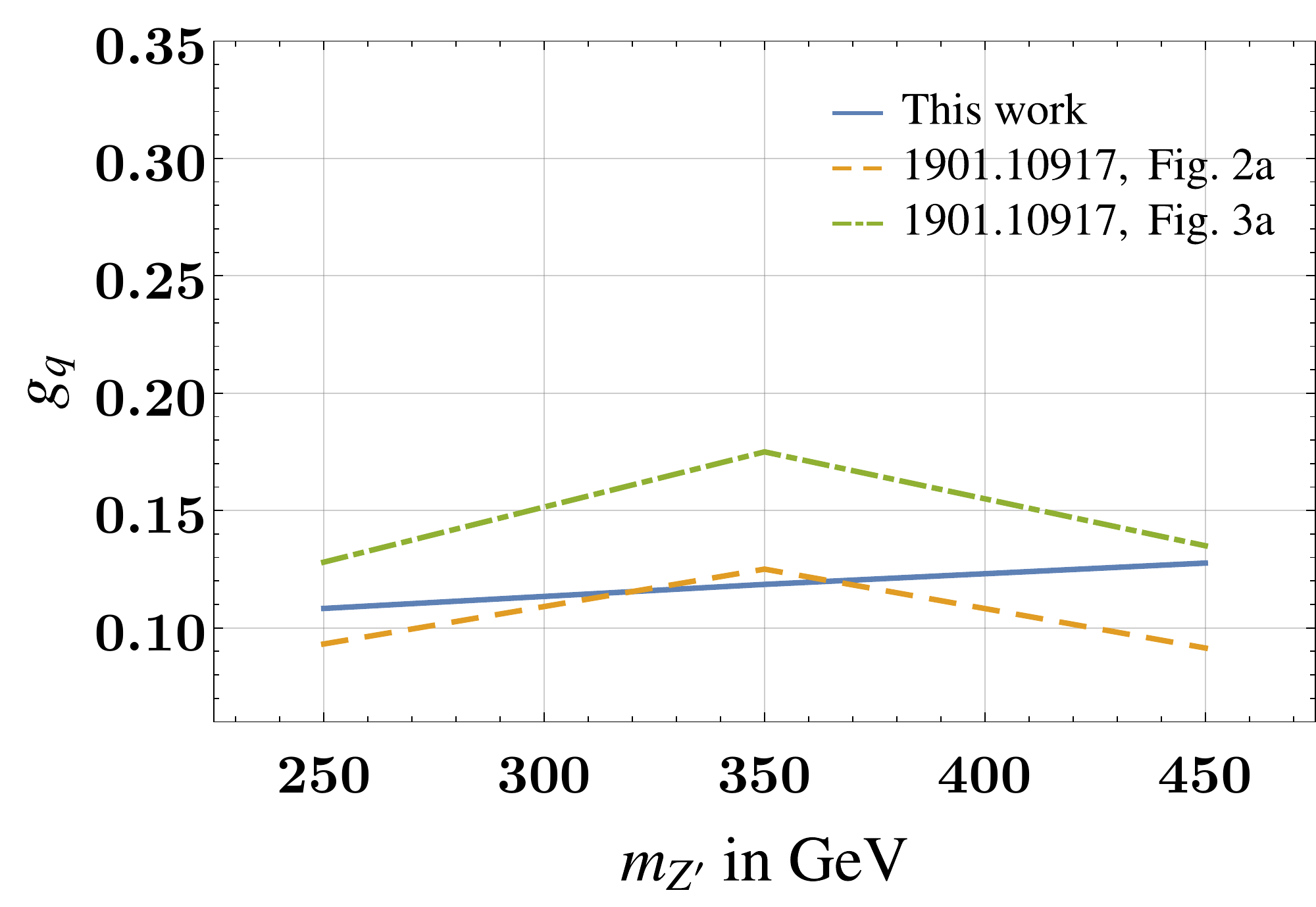}
    \caption[Comparison with the experimental analysis.]{Comparison of the acceptance (left) and of the limit on the coupling (right)
            between our analysis and the experimental analysis from Ref.~\protect\cite{Aaboud:2019zxd}
            (1901.10917).}
    \label{fig:acceptance and couplings bound}
\end{figure}

Here we cross-check our analysis procedure which we used to recast the results of Ref.~\cite{Aaboud:2019zxd},
cf. Section~\ref{sec:qb qb}.
For this we apply our code to the $Z'$ model that has been analysed in the very same reference.
This enables a direct comparison between our analysis and the experimental one.
The data for the process $pp\rightarrow \gamma Z'(\rightarrow jj)$ is generated
in \texttt{MadGraph5\_aMC@NLO} using the vector-leptoquark model file of Ref.~\cite{Baker:2019sli}.
The acceptance of our procedure is shown together with the acceptance of the
experimental analysis, that can be found in the auxiliary materials of Ref.~\cite{Aaboud:2019zxd},
in Fig.~\ref{fig:acceptance and couplings bound}.
Further, we determine the limits on the coupling as described in Section~\ref{sec:qb qb}.
The result is shown in Fig.~\ref{fig:acceptance and couplings bound} where we overlaid
the results of Ref.~\cite{Aaboud:2019zxd}.
Both results show a good agreement with the more sophisticated experimental
analysis.

\section{Validation of \texttt{PYTHIA} code used for $qb \to H^\pm \to tb$ analysis}
\label{validation2}

In order to validate our \texttt{PYTHIA} code that we used for the $qb \to H^\pm \to tb$ analysis,  we have used the same code to reproduce the number of simulated SM background events after cuts in various channels as obtained by the experimental collaborations in the Ref.~\cite{ATLAS:2015nkq} (ATLAS) and Ref.~\cite{CMS:2012jea} (CMS). 

\subsection{Validation with arXiv:1209.4397 (CMS)}
The CMS analysis was dedicated to the production of heavy resonances decaying into top-antitop quark pairs using $5.0$ fb$^{1}$ at $\sqrt{s} = 7$ TeV.  SM background events from $t \overline{t}$ production were generated using \texttt{MADGRAPH 5.1.1}, \texttt{PYTHIA 6.4.24} (generating $pp \to t \overline{t}$ with up to two additional jets)  and \texttt{POWHEG} event generators using CTEQ6L parton distribution functions of the proton. On these events selection cuts were imposed by requiring exactly one isolated muon with $p^l_T>30\GeV$ and $|\eta^l|<2.1$ and at least three jets with transverse momentum $p_T^j>50\GeV$, $|\eta|<2.5$, where the leading leading jet hast at least $p_T>70\GeV$. The anti-$k_t$ algorithm with $R=0.5$ was used for jet clustering, where candidates identified as leptons were excluded from jet clustering. Moreover, missing transverse energy of at least 20 GeV were required, and only tracks with 
transverse momentum above 0.5 GeV were taken into account. Different signal categories were obtained by carying the number of required jets and $b$-tagged jets. The cross sections have been normalized to the NLO values given in Ref.~\cite{CMS:2012jea} ($\sigma_{t \overline{t}}^{7 \TeV} = 157.5 \, {\rm pb}$), corresponding to a $K$-factor of 1.51. 

In Tab.~\ref{CMSvalidation} we display the numbers of events which passed all cuts in the various signal categories,  where e.g.``$\ge 4j, \, 1b$'' means at least four jets of which there is one $b$-tagged jet. Within statistical errors we find excellent agreement with the simulated events obtained by the CMS collaboration. 
 \begin{table}[h]
\center
     \begin{tabular}{ccccc}
    \toprule
                          & $3j, \, \geq1b$  & $\geq4j, \, 0b$  & $\geq4j, \, 1b$ & $\geq4j, \, \geq2b$  \\
    \midrule 
      Our analysis        & $7657\pm 1245$   & $2214\pm668$     & $8261\pm1294$   & $7657\pm1245$  \\
       \midrule
      Ref.~\cite{CMS:2012jea}  & $5612$      & $2988$           & $7802$          & $6093$         \\
    \bottomrule
    \end{tabular}
    \caption{Expected number of events for $t \overline{t}$ production at $\sqrt{s} = 7\,$TeV and $5.0\,$fb$^{-1}$ obtained with our analysis code and by the CMS collaboration in Ref.~\cite{CMS:2012jea}.  For our analysis we only show the statistical errors from the limited MC samples we used, whereas the errors in the numbers of Ref.~\cite{CMS:2012jea} are dominated by systematic uncertainties, which amount to roughly 10\%. \label{CMSvalidation}}
\end{table}

\subsection{Validation with arXiv:1512.03704 (ATLAS)}
The ATLAS analysis was dedicated to the search for charged Higgs bosons produced in association with a top quark $g b \to t H^\pm$ decaying to $tb$ using $20.3$\,fb$^{-1}$ at $\sqrt{s} = 8$ TeV. SM background events from $t \overline{t}$ production were simulated  with \texttt{Powheg-Box v2.0}, using the CT10 PDF set. These events were interfaced to \texttt{Pythia v6.425}, with the Perugia P2011C tune for the underlying event. Events were further selected by requiring exactly one isolated lepton, satisfying $E_T>25\GeV, |\eta|<2.47$ (electrons) and $p_T>25 \GeV, |\eta|<2.5$ (muons), and at least 4 jets with transverse momentum $p_T^j>25\GeV$ and pseudo-rapidity  $|\eta^j|<2.5$. The anti-$k_t$ algorithm with $R=0.4$ was used for jet clustering, where candidates identified as leptons were excluded from jet clustering. Moreover, missing transverse energy of at least 20 GeV were required, and only tracks with transverse momentum above 0.5 GeV were taken into account. The $b$-tagging algorithm has 70\% efficiency to tag a $b$-quark jet, with
a light-jet mistag rate of 1\% and a $c$-jet mistag rate of 20\%. The $t \overline{t}$  cross sections have been normalized to the NNLO values given in Ref.~\cite{ATLAS:2015nkq} ($\sigma_{t \overline{t}}^{8 \TeV} = 253 \, {\rm pb}$), corresponding to a $K$-factor of 1.72. 

In Tab.~\ref{ATLASvalidation} we display the numbers of events which passed all cuts in the various signal categories,  where e.g.``$\ge 4j, \, 3b$'' means at least four jets of which there are three $b$-tagged jet. Within statistical errors we find excellent agreement with the simulated events obtained by the ATLAS collaboration.
\begin{table}[ht]
\center
     \begin{tabular}{ccccccc}
    \toprule
                                & $4j, \, 2b$     & $5j, \, 2b$     & $\geq6j, \, 2b$  & $4j, \, \geq3b$  & $\geq5j, \, \geq3b$ \\
    \midrule 
      Our analysis              & $71508\pm3142$ & $41654\pm 2374$ & $28631\pm 1990$  & $5063\pm 838$   & $8624\pm 1082$    \\
    \midrule
      Ref.~\cite{ATLAS:2015nkq} & $87220\pm13740$ & $44750\pm10830$  & $24490\pm8420$   & $7700\pm1780$    & $9700\pm 3800$     \\
    \bottomrule
    \end{tabular}
    \caption{$\bm{t\overline{t}js}$: Expected number of events for $t \overline{t}$ production at $\sqrt{s} = 8\,$TeV and $20.3\,$fb$^{-1}$ obtained with our analysis code and by the ATLAS collaboration in Ref.~\cite{ATLAS:2015nkq}. For our analysis we only show the statistical errors from the limited MC samples we used, whereas the errors Ref.~\cite{ATLAS:2015nkq} are dominated by systematic uncertainties.  \label{ATLASvalidation}}
\end{table}

\section{BDT analysis of $qb \to H^\pm \to tb$ signature}
\label{BDT}
In order to maximise the reach of the search, we perform a multivariate
analysis by employing a boosted-decision-tree (BDT).
For this, we use an AdaBoost classifier as implemented in the \texttt{scikit-learn}
library in Python.
For Higgs masses below (equal and above) $800\GeV$, we train the classifier using $10$
trees and a learning rate of $0.4$ with each tree having a maximal depth of $14$ ($8$).

\begin{figure}[t]
\includegraphics[width=0.5\textwidth]{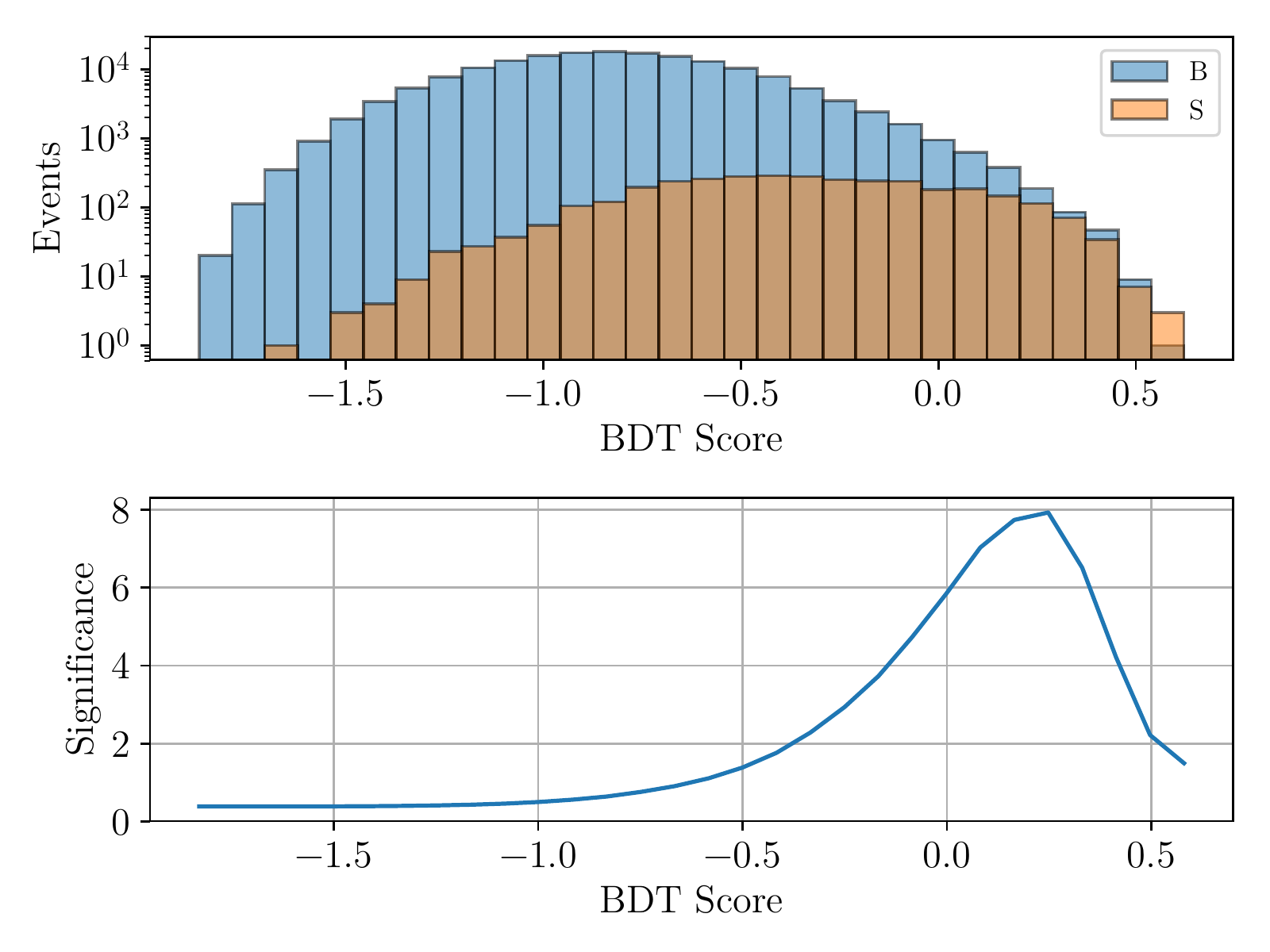}
\includegraphics[width=0.5\textwidth]{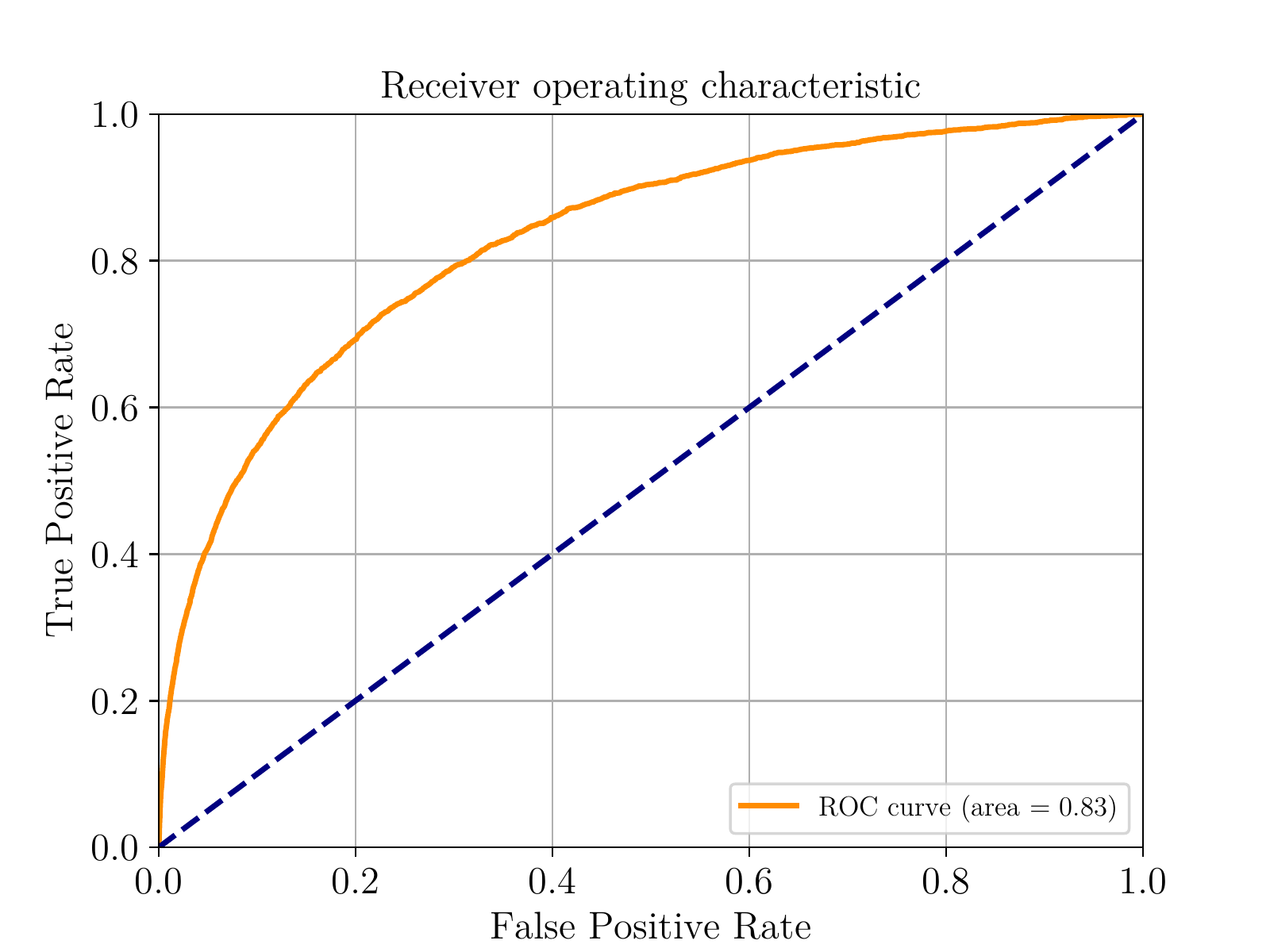}
\caption{$\mathcal{L}=140\,\mathrm{fb}^{-1}$. \textit{Left}: Significance $Z \approx S/\sqrt{S + B + (\epsilon B)^2}$ for
$\epsilon = 5\%$ as a function of the BDT cut.
    \textit{Right}: Receiver operating curve.}
\label{fig:BDT}
\end{figure}

The BDT is trained on $10^5$ unweighted $t\overline{t}$ background events and
$10^4$ unweighted signal events which have passed the selection region defined as
\begin{align}
    p_T^j & > 20\GeV \,, &p_T^l& > 30\GeV \,,
    &H_T& > 350 \GeV \,, &E_T^\mathrm{miss}&> 35\GeV \,,
    &|\eta|& < 2.5 \,,
\end{align}
with exactly one lepton and at least three $b$-jets, cf. Section~\ref{qbtb}
The variables included in the training are
\begin{itemize}
    \item $p^b_{T,i}$ of the three leading $b$-jets, $i=1,2,3$
    \item $p^l_T$ of lepton
    \item scalar sum of the $p_T$ of the three $b$-jets and lepton
    \item missing energy $E_T^\mathrm{miss}$
    \item $\Delta R$ between the three leading $b$-jets and the lepton,
        denoted as $R^b_{ij}$ and $R_{li}$ for the distance between the
        $i$th and $j$th $b$-jet, and the $i$th $b$-jet and lepton,
        respectively
    \item pseudo-rapidity of the three leading $b$-jets $\eta_i^b$ and the lepton $\eta^l$
    \item number of jets $n_\mathrm{jets}$
    \item number of $b$-jets  $n_{b\text{-jets}}$
\end{itemize}
Note that the BDT is only trained on the signal events and the dominant
$t\overline{t}$ background, however, the performance evaluated
with a separate dataset for a luminosity of $\mathcal{L} = 140\,\mathrm{fb}^{-1}$
takes into account all processes listed in Tab.~\ref{comparison}.
We show the results exemplary for the $450\GeV$ charged-Higgs mass point
with $g_{cb} = 0.4$ and $g_{tb} = 0.6$ and a systematic uncertainty of
$\epsilon = 5\%$.
The receiver operating curve showing the true positive over
false positive rate is shown on the right plot of Fig.~\ref{fig:BDT} while
the BDT output is shown on left plot.
The latter is treated as an additional cut variable,
thus cutting at a BDT output of $0.25$ yields the maximal significance of $\approx 7.9$.

\begin{figure}[t]
    \centering
    \includegraphics[width=0.85\textwidth]{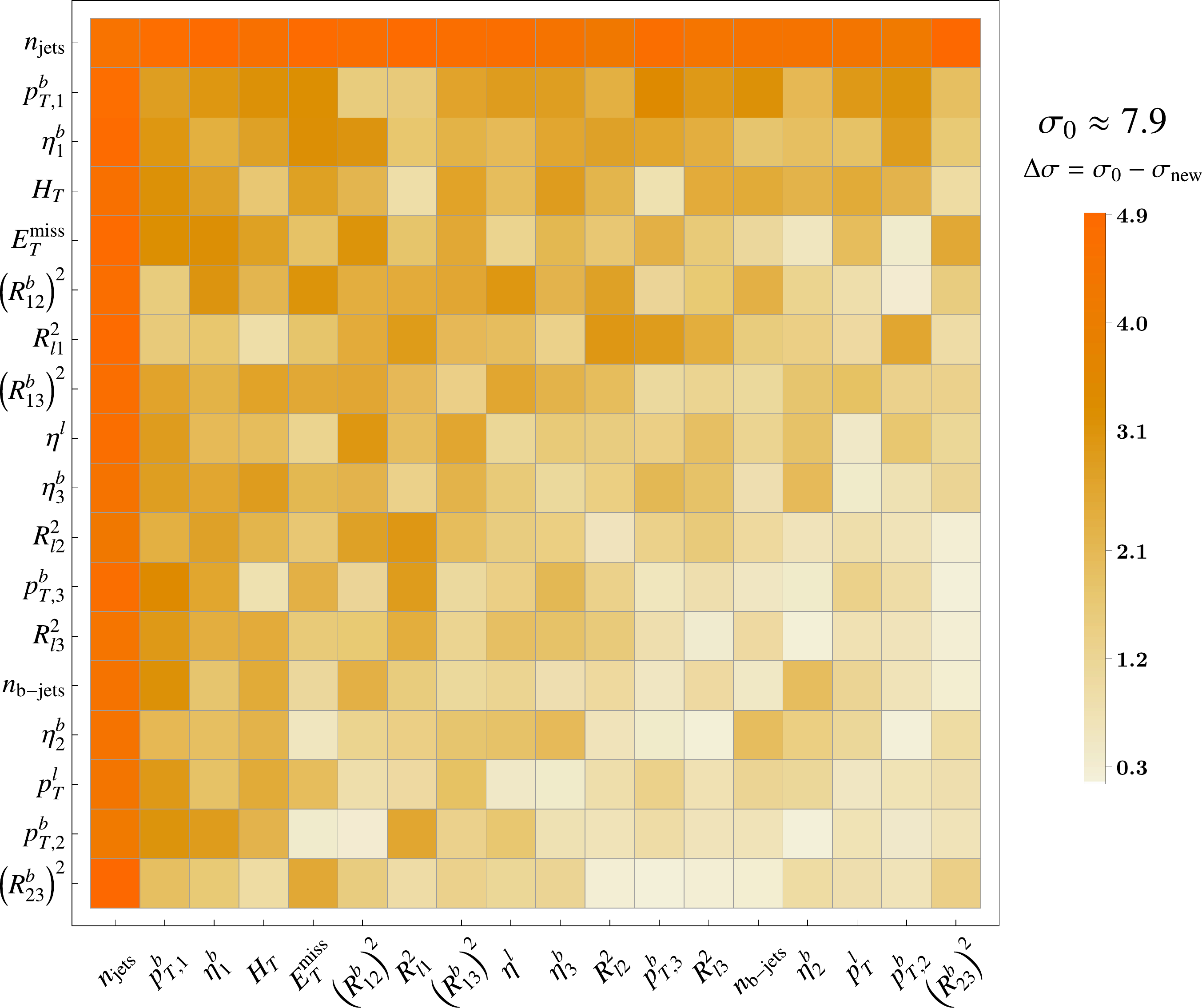}
    \caption{Change in the significance $\Delta Z$ when re-evaluating the BDT
        after performing the training without the variables listed on the axes. The diagonal
        elements correspond to dropping one training variable. The matrix is by construction
        symmetric. The significance using the whole set of variables is $\sigma_0 \approx 7.9$.}
    \label{fig:impact BDT variables}
\end{figure}

To further determine the BDT variables with the most impact,
we drop up to two training variables and retrain the BDT without this
information. The resulting BDT is evaluated on the same
$\mathcal{L} = 140\,\mathrm{fb}^{-1}$ dataset as before. The resulting drop
in the significance is shown in Fig.~\ref{fig:impact BDT variables}.


\clearpage
\bibliographystyle{JHEP} 
\bibliography{Bibliography}

\end{document}